\documentclass[12pt]{amsart}
\usepackage[utf8x]{inputenc}
\usepackage{fullpage}
\usepackage{float,rotating}
\usepackage{caption}
\usepackage{mathtools}
\usepackage{parskip}
\usepackage[dvipsnames,table]{xcolor}
\usepackage{subfig}
\usepackage{breqn}
\usepackage[multiple]{footmisc}

\definecolor{cadetgrey}{rgb}{0.57, 0.64, 0.69}
\PassOptionsToPackage{hyphens}{url}
\usepackage[colorlinks = true,
            linkcolor = blue,
            urlcolor  = blue,
            citecolor = blue,
            anchorcolor = blue]{hyperref}
\usepackage{multirow,multicol, makecell, booktabs}

\colorlet{shadecolor}{gray!90}
\usepackage{bbm}
\usepackage{graphicx}
\usepackage{longtable}
\usepackage{booktabs,array,multirow}
\usepackage{amsfonts,amsmath,amssymb}
\usepackage{amsbsy}
\usepackage{amstext}
\usepackage{amsthm}
\usepackage{amssymb}

\newif\iflatexml\latexmlfalse
\AtBeginDocument{\DeclareGraphicsExtensions{.pdf,.PDF,.eps,.EPS,.png,.PNG,.tif,.TIF,.jpg,.JPG,.jpeg,.JPEG}}
\usepackage{url}

\usepackage[authoryear]{natbib}
\usepackage{graphicx}

\usepackage{setspace}
\providecommand{\algorithmname}{Method}
\providecommand{\corollaryname}{Corollary}
\providecommand{\propositionname}{Estimation}

\theoremstyle{plain}

\theoremstyle{plain}

\theoremstyle{plain}

\usepackage{ulem}
\usepackage{subfloat}
\usepackage{pdflscape}
\usepackage{everypage}
\usepackage{lipsum}

\usepackage{accents}

\newcommand{\Lpagenumber}{\ifdim\textwidth=\linewidth\else\bgroup
  \dimendef\margin=0 %use \margin instead of \dimen0
  \ifodd\value{page}\margin=\oddsidemargin
  \else\margin=\evensidemargin
  \fi
  \raisebox{\dimexpr -\topmargin-\headheight-\headsep-0.5\linewidth}[0pt][0pt]{%
    \rlap{\hspace{\dimexpr \margin+\textheight+\footskip}%
    \llap{\rotatebox{90}{\thepage}}}}%
\egroup\fi}
\AddEverypageHook{\Lpagenumber}%
\usepackage{tikz}
  \usetikzlibrary{shapes,arrows,fit,calc,positioning}
\tikzset{
  basic/.style  = {draw, text width=4cm, rectangle},
    root2/.style   = {basic, rounded corners=2pt, thin, align=center,text width=16em},
  root/.style   = {basic, rounded corners=6pt, thin, align=center,text width=15em},
  level 2/.style = {basic, rounded corners=6pt, thin, align=center,
                   text width=14em},
  level 3/.style = {basic, thin, align=left, rounded corners=2pt, text width=17em, scale=0.7, draw=white},
level 4/.style = {basic, rounded corners=6pt, thin, align=left,
                   text width=11em, scale=0.9},
  level 5/.style = {basic, thin, align=left, rounded corners=2pt, text width=6.4em, scale=0.6},
  level 6/.style = {basic, rounded corners=6pt, thin, align=center,
                   text width=6em},    
  level 7/.style = {basic, rounded corners=6pt, thin, align=center,
                   text width=6em}                      
}
  \tikzset{line/.style={draw, thick, -latex'}}
  
\usepackage{array}
\newcolumntype{P}[1]{>{\raggedleft\arraybackslash}p{#1}} 
\newcolumntype{C}[1]{>{\centering\arraybackslash}p{#1}} 
\usepackage{cleveref}
\usepackage{lscape}
\usepackage{tabulary}
\usepackage{scalerel,stackengine}
\stackMath
\newcommand\reallywidehat[1]{%
\savestack{\tmpbox}{\stretchto{%
  \scaleto{%
    \scalerel*[\widthof{\ensuremath{#1}}]{\kern-.6pt\bigwedge\kern-.6pt}%
    {\rule[-\textheight/2]{1ex}{\textheight}}%WIDTH-LIMITED BIG WEDGE
  }{\textheight}% 
}{0.5ex}}%
\stackon[1pt]{#1}{\tmpbox}%
}
\usepackage{fancyhdr}
\usepackage[yyyymmdd,hhmmss]{datetime}
\definecolor{cadetgrey}{rgb}{0.57, 0.64, 0.69}
\usepackage{epigraph}
\begin{document}
\title{Quantifying Uncertainties in estimates \\of income and wealth inequality  \\ \vspace{1cm}
\author{Marta Boczo\'n \\\vspace{0.1cm}
\tiny \normalfont{October, 2020}\\
\tiny \normalfont{\tiny \href{https://martaboczon.com}{(Click here for the latest version)}}}}

\thanks{\textit{Author:} Boczo\'n: Department of Economics, University of Pittsburgh, 230 South Bouquet Street, Wesley W. Posvar Hall, Pittsburgh PA, 15213. E-mail: martaboczon@pitt.edu.  Foremost, I would like to express my sincere gratitude to my supervisor Jean-Fran\c{c}ois Richard,  to whom I would like to dedicate this project, for his continuous support throughout my Ph.D. studies and particularly for his dedication, encouragement, and immense knowledge. He has been not only my primary advisor but also my mentor.  Besides my supervisor, I would like to thank the rest of my thesis committee: Arie Beresteanu, Roman Liesenfeld,  and Alistair Wilson for their motivation and insightful comments. My completion of this project could not have been possible without the support of Jesse Bricker, John Czajka, Daniel Feenberg,  Kevin Moore, William Peterman, and Allison Schertzer. For helpful feedback, I also thank Claudia Sahm, Janine Carlock, and the University of Pittsburgh Writing Center. Lastly, I am grateful to the Federal Reserve Board of Governors Dissertation Internship program for research support. Remaining errors are my sole responsibility.}
\begin{abstract}\begin{singlespace}
I measure the uncertainty affecting estimates of economic inequality in the US and investigate how accounting for properly estimated standard errors can affect the results of empirical and structural macroeconomic studies. In my analysis, I rely upon two data sets: the Survey of Consumer Finances (SCF), which is a triennial survey of household financial condition, and the Individual Tax Model Public Use File (PUF), an annual sample of individual income tax returns. While  focusing on the six income and wealth shares of the top 10 to the top 0.01 percent between 1988 and 2018, my results suggest that ignoring uncertainties in estimated wealth and income shares can lead to erroneous conclusions about the current state of the economy and, therefore, lead to inaccurate predictions and ineffective policy recommendations. My analysis suggests that for the six top-decile income shares under consideration, the PUF estimates are considerably better than those constructed using the SCF; for wealth shares of the top 10 to the top 0.5 percent, the SCF estimates appear to be more reliable than the PUF estimates; finally, for the two most granular wealth shares, the top 0.1 and 0.01 percent, both data sets present non-trivial challenges that cannot be readily addressed. 
 \end{singlespace}
\end{abstract}
\maketitle

\section{Introduction}
As of 2018, more than 40 percent of income was earned by the top 10 percent, and more than 30 percent by the top 5 percent. In relation to wealth, more than 75 percent was owned by the top 10 percent, and more than 65 percent by the top 5 percent. According to \citet{saez_2017_income_and_wealth_inequality}, the last time we observed comparably high levels of top income and wealth inequality was in the years leading to the 1929--1933 Great Depression. 

Inequality endangers the economy in a number of ways:  it  threatens the integrity of economic systems and the impartiality of political institutions, and eventually can lead to a rise of extremism or even oligarchies.  So far, income and wealth inequality in the US has been linked to declining 
trust in political institutions, low voter turnout, political polarization, declining life expectancy, and a rise in obesity, mental illness, homicide, teenage pregnancy, and substance abuse. 
For example, \citet{saez_2019_the_triumph} find that rich Americans live almost 15 years longer than poor ones. This represents a gap in life expectancy comparable to that between the US and Nigeria. Moreover, rising income and wealth concentration in the US endangers equal distribution of economic resources around the world. In particular, between 1970 and 1992 an increase in the number of ``globally rich'' in the US, defined as those with more than twenty times the mean world income, accounted for half of the worldwide increase, making ``a perceptible difference to the world distribution'' \citep{atkinson_2011_topincomes}.

While numerous studies, including \citet{income_2003_piketty}, \citet{atkinson_2011_topincomes}, \citet{bricker_2016_measuring}, \citet{bricker_2018_how_much}, and \citet{saez_2016_wealthinequality, seaz_2020_revising_after_the_revisionsists} estimate income and wealth inequality, few examine the statistical uncertainty around their estimates. However, this is of great importance, since such uncertainties could lead to erroneous conclusions 
about the current state of the economy and, therefore, result in inaccurate predictions and ineffective policy recommendations. 

During the past two years, economists and politicians have discussed various measures aimed at combating inequality: instituting a wealth tax on millionaires, raising the top income tax rate, reducing exemptions and increasing tax rates on large estates. 
However, whether such policies would prove effective at closing the gap between rich and poor depends primarily on our ability to produce statistically 
accurate estimates of income and wealth inequality. Otherwise, the government might collect either too little in tax revenue, 
unable to provide the poor with adequate government-funded child care and paid-leave, or too much, causing a sudden and sharp decline in economic growth. 

Moreover, since income and wealth inequality has been at the center of attention during the 2020 Democratic Party presidential primaries, 
it is imperative to provide the general public with an idea of the accuracy of these estimates. Otherwise, the public could easily be
misled to either under- or overestimate the severity of the ongoing crisis linked to rising inequality. This, in turn, may result in voters misconstruing the effectiveness of current policies
and lead them to express support for more conservative proposals. 

In this paper, I estimate the uncertainties  in estimates of the six income and wealth shares of 
the top 10 to the top 0.01 percent (the American upper middle and upper classes)  and assess their impact on both empirical and
structural macroeconomic studies. To do this, I first investigate which data set proves most credible for studying income and wealth concentration. Second, I analyze whether my results regarding the magnitudes and trends in top income and wealth inequality support or contradict those published in the related literature. Finally, I examine to what extent uncertainties in calibration targets affect outcomes of structural macroeconomic modeling, in the context of a random growth model of income.

The present paper contributes to the existing literature on economic inequality in five main ways. First, it adds to our understanding of the types of error that are most prevalent in estimates of income and wealth inequality. Second, it provides a cost-benefit analysis of studying economic inequality using survey data (with a wide range of both financial and nonfinancial variables but a small number of observations) versus administrative tax records (with a large number of observations but no information on taxpayers' demographic or socio-economic characteristics). Third, it introduces a novel approach to estimating sampling error using administrative tax data, with a focus on the context of economic inequality.  Fourth, it is the first research project to estimate the long-term dynamics in economic inequality while accounting for uncertainties in the constructed estimates of top income and wealth shares. Finally, it discusses the consequences of utilizing error-prone data for structural analysis, including data tracking and projecting.

In this paper, I define income as gross income comprising all income items except for capital gains and wealth as all assets less all liabilities. For both the empirical and structural analysis, I use two data sets: the Survey of Consumer Finances (SCF)---a triennial survey of US household financial condition---and the Individual Tax Model Public Use File (PUF)---an annual sample of US individual income tax returns. The SCF survey data range from 1988 to 2018, and the PUF administrative data range from 1991 to 2012 (the 2018 SCF and the 2012 PUF are the latest available data sets at the time of writing). Consequently, my analysis focuses on the over twenty-year long period that follows the Tax Reform Act of 1986, which lowered federal income tax rates and, in particular, reduced the top tax rate from 50 to 28 percent.

In order to determine which data source is more reliable for studying top income and wealth inequality, I compare the SCF survey data and the PUF administrative data with respect to a number of criteria, such as the total and weighted number of observations available for the estimation and the size of relative standard errors. For studying long-term dynamics in income and wealth inequality, I use weighted least squares with weights defined as reciprocals of squared standard errors. In addition to long-term trends, I examine how income and wealth concentration changed between the onset and the aftermath of the 2007--2009 Great Recessions. 

In addition to accounting for data-driven errors in empirical analysis of top-decile income and wealth shares, I investigate how data deficiencies affect outcomes of structural macroeconomic models. I consider the augmented random growth model of income proposed by \citet{gabaix_2016_dynamics} and use Monte-Carlo simulation techniques to analyze how errors in inputs to this model impact the precision of  the model's outcomes. Specifically, for each of the two data sets under consideration, I calibrate \citet{gabaix_2016_dynamics}'s model multiple times, each time using a different value randomly drawn from a confidence interval constructed around the point estimate of the model's calibration target. Then, by averaging over the range of generated model outcomes, I determine the extent to which the model's outputs are affected by the uncertainty in the model's inputs.

My empirical analysis of top income inequality suggests that estimates constructed using the administrative data are considerably better than those constructed using the survey data. One of the advantages of using the PUF is its higher data frequency and the fact that sampling error in the estimated income shares is on average \textit{five} times smaller than in the SCF. While these data features are not critical when examining long-term dynamics of income inequality, they become a decisive factor when choosing between the SCF and the PUF in a study that analyzes short-time horizons and year-to-year changes. Moreover, the small number of observations above the 99.9 and 99.99 income fractiles in the SCF  makes the SCF estimates of the two most granular income shares of the top 0.1 and 0.01 percent extremely volatile and, thus, uninformative. 

In relation to wealth inequality, my empirical results indicate that neither the survey data nor the administrative data can be used without caution. For the less granular wealth shares of the top 10 to the top 0.5 percent, I find the SCF more reliable than the PUF. In the SCF, respondents are \textit{asked} about their asset and liability holdings, whereas in the PUF, the wealth of every individual in the sample is \textit{estimated} from their reported income using a capitalization model. Since such models are heavily dependent on numerous (and often arbitrary) assumptions imposed on assets' rates of return, so are the resulting estimates of top wealth shares. Therefore, even though the SCF estimates have larger sampling errors than those constructed using the PUF, they are free from non-trivial and yet-to-be-fully-determined modeling errors arising in the process of inferring wealth from income. Lastly, regarding wealth shares of the top 0.1 and 0.01 percent, I find that both the SCF and the PUF present difficult-to-overcome challenges (an insufficient number of observations and modeling errors, respectively) that result in highly unreliable estimates of the far right tail of wealth distribution of the top 0.1 and 0.01 percent. 

In addition to identifying strengths and weaknesses of survey and administrative data in studying income and wealth concentration, this paper adds new insight to the ongoing debate regarding the magnitudes and trends in top income and wealth inequality. For income, my results confirm those from the related literature, indicating a statistically significant increase in income concentration between the early 1990s and the early 2010s, and suggest comparable levels and trends in the estimated income shares within the top 10 percent. 

For wealth, since this paper finds the SCF a more reliable data source than the PUF, it portrays a different picture of top wealth inequality than the most widely-cited studies on wealth concentration estimated using administrative tax-level data  \citep[e.g.,][]{saez_2016_wealthinequality}. Specifically, while my study does suggest a statistically significant increase in the wealth shares of the top 10, 5, and 1 percent between the early 1990s and the early 2010s, the estimated trend lines are more modest than in \cite{saez_2016_wealthinequality}. Second, I do not find evidence of rising wealth inequality \textit{within} the top ten percent. Unlike \cite{saez_2016_wealthinequality} who observe a larger increase in the wealth shares of the top 1 percent than in the wealth shares of the top 10 percent, the weighted linear regression analysis of the SCF point estimates suggests the opposite, casting doubts on a presumption that an observed rise in wealth inequality is driven solely by the far right tail of wealth distribution.  Furthermore,  this paper does not support the authors' widely-cited conclusion regarding a 100 percent increase in the wealth shares of the top 0.1 percent between 1991 and 2012, a finding that has drawn substantial media coverage and major interest from politicians and policy makers.

My third set of results pertains to the consequences of modeling inequality using error-prone data. In relation to the \citet{gabaix_2016_dynamics}'s random growth model of income, I find that having precise estimates of calibration
targets is critical for producing precise outcomes of structural analysis. This is the case since errors in calibration targets are carried over through the model and come to affect all outcomes of interest. Specifically, I find that the model calibrated to administrative tax data projects income shares of the top 1 percent in 2050 to be equal to 22.5 percent, with only negligible levels of uncertainty attributable to the data. On the other hand, the model calibrated to survey data is much less precise regarding the 2050 projection, with a 95 confidence interval ranging from 19 to 29 percent. Therefore, by relying upon administrative tax data for the model's calibration, as opposed to survey data, one can reduce the uncertainty in the model's outcome of interest by a factor of ten.

My paper is organized as follows. In Section \ref{sec:lit_review}, I discuss related literature. In Section \ref{sec:tse}, I provide a brief description of sources of error in survey and administrative data. In Section \ref{sec:Income inequality in survey data}, I characterize the main features of the SCF and describe the estimation procedure of the SCF standard error. In Section \ref{sec:income_inequality_in_administrative_data}, which follows the same format as Section \ref{sec:Income inequality in survey data}, I first provide a brief description of the PUF and next, characterize the estimation of the PUF standard error.
In Section \ref{sec:income_inequality}, I define the concepts of income and wealth and discuss the estimation procedure of top-decile income and wealth shares. In Sections \ref{sec:empirical_results_income} and \ref{sec:wealth_empirical_results}, I analyze the main empirical results centered around income and wealth inequality, respectively. In Section \ref{sec:structural_exercise}, I discuss the key outcomes of the structural analysis. Section \ref{sec:conclusions} concludes. Online supplementary material with additional
results and detailed discussions supporting my conclusions is available on \href{https://martaboczon.com}{https://martaboczon.com}.

\section{Literature review}
\label{sec:lit_review}
This paper contributes to four main strands of economic literature: economic inequality, survey statistics, SCF survey design, and PUF sample design.

First, since one of my objectives is to quantify uncertainties in estimates of income and wealth inequality within the top 10 percent, this paper contributes to the growing body of literature on income and wealth inequality in the US. Specifically, it is closely-related to the work of  \citet{income_2003_piketty}, where the authors rely upon tax returns statistics and micro-level data on individual-income tax returns in order to construct homogeneous series of top-decile income shares  between 1913 and 1998. Another important reference the research is related to is \citet{atkinson_2011_topincomes}, which utilizes individual-income tax return statistics for numerous income brackets in order to provide a comprehensive overview and comparative analysis of historical and current trends in top income shares for multiple countries around the globe. 

In addition, the current paper adds valuable insights to the literature on wealth inequality. In particular, it builds on \citet{saez_2016_wealthinequality} and \citet{bricker_2018_how_much}, in which the authors rely upon micro-level data and aggregate statistics published in the Financial Accounts of the United States in order to estimate the distribution of wealth using a capitalization model.  Moreover, it is indirectly related to \citet{kopczuk_2004_top_wealth_shares}, who 
estimate wealth concentration using estate tax return data, in which individual estates are weighted by the inverse probability of death.

Since this paper analyzes various data-driven errors in surveys and other sample data, it constitutes a direct application of an important statistical concept related to the Total Survey Error (TSE) paradigm. Even though, TSE has been thoroughly discussed in the literature on survey statistics \citep[see, e.g.][]{biemer_2003_introduction, groves_2009_survey}, it remains largely ignored in economics. Therefore, this paper constitutes an example of how established and widely applied statistical concepts can benefit economic research.

In addition to adding to economic inequality and survey statistics literature, the present paper contributes to the literature on the SCF survey design  \citep[see, e.g.][]{kennickell_1997_consistent, kennickell_1998_multiple, kennickell_2000_revisions, kennickell_2008_role}. Specifically, except for research conducted by the Board of Governors of the Federal Reserve System, it is the first academic paper to consider data-driven errors in any macroeconomic estimates constructed using the SCF survey data.

Lastly, this paper contributes to the literature on the PUF sampling design \citep[see][]{czajka_2014_assessment, bryant_2014_design}. Specifically, it proposes a bootstrapping technique that allows data users to estimate the PUF sampling error for any quantity of interest. As such, it provides an illustrative example of how the information regarding a complex  sample selection process can be incorporated into an economic analysis. 

\section{Total Survey Error}
\label{sec:tse}
Since one of my primary objectives is to identify and quantify sources of data-driven errors in the estimates of income and wealth inequality, this paper is centered around the TSE paradigm---an umbrella term for a variety of error sources in survey data. Even though TSE pertains primarily to errors in surveys, it is also applicable to data consisting of administrative records. This is the case since administrative data are affected by the same sources of error as survey data. Specifically, as with any other sample, administrative data are subject to sampling error caused by drawing a sample rather than conducting a complete census. Moreover, the data are prone to nonsampling errors, which comprise all other sources of error arising in the process of designing, collecting, processing, and analyzing of sample data. 

TSE consists  of two main components: sampling error and nonsampling error, which can be further divided into specification, frame, nonresponse, measurement, and processing errors, all of which I briefly characterize in the remainder of the present section.\footnote{In this paper, without loss of generality I rely upon a TSE decomposition from \cite{biemer_2003_introduction}. Alternative decomposition can be found, for example, in \citet{groves_2009_survey}.} One of the advantages of decomposing TSE is that it allows me to  differentiate between sources of error, and consequently, address them individually. Specifically, in the present paper, I estimate sampling and nonresponse errors in the SCF and sampling error in the PUF. As such, I do not account for all parts of TSE (which is beyond the scope of the present paper). Instead, I provide qualitative evidence on which components of TSE can be considered marginal for this particular analysis and which are likely to be non-negligible and therefore, to be accounted for in a follow-up research project.

Specification error occurs ``when the concept implied by the survey question
and the concept that should be measured in the survey differ'' \citep{biemer_2003_introduction}.  This often results from  misunderstandings between the  different parties involved in the survey process such as researchers, data analysts, survey sponsors, questionnaire designers, and others. 

The other sources of nonsampling error are frame and processing errors. The former relates to the process of constructing, maintaining, and using the sampling frame for selecting the sample, whereas the latter occurs in data editing, coding, entry of survey responses, assignment of survey weights, tabulation and other data arrangements.\footnote{A classic frame error occurred in a public opinion poll designed to predict the result of the 1936 presidential election between Alfred Landon and Franklin D. Roosevelt. Since the sample frame was heavily over-represented by individuals who identified as Democrats (phone owners, magazine subscribers, members of professional associations), the difference between the poll's
prediction and the election's
result was equal to 19 percentage points and as such, constitutes the largest error ever recorded in a major public opinion poll.} 

Nonresponse  encompasses unit nonresponse, item nonresponse, and incomplete response, and is considered ``a fairly general source of error'' \citep{biemer_2003_introduction}. A unit nonresponse occurs when a sampling unit  does not participate in the survey; an item nonreponse when a participating unit leaves a blank answer to a specific survey question; and an incomplete response when the answer provided to a typically open-ended question is either incomplete or inadequate. 
 
The fifth and final source of nonsampling error, measurement error,  is considered ``the most damaging source of error'' \citep{biemer_2003_introduction}. It includes errors arising from respondents and interviewers, in addition to other factors such as the design of the questionnaire, mode of data collection, information system, and interview setting.

In summary, any estimator constructed using survey or sample data is subject to a variety of sampling and nonsampling errors. Therefore, an important question is that of the extent to which these errors can be reliably accounted for when estimating unknown population parameters (such as top income and wealth shares) using self-reported survey data and administrative records.

\section{Uncertainties in survey data on consumer finances}
\label{sec:Income inequality in survey data}
The SCF is a triennial survey of household finances sponsored by the Board  in cooperation with the Statistics of Income (SOI) Division of the Internal Revenue Service (IRS).\footnote{Until 1988 the survey data were collected by the Survey Research Center at the University of Michigan. Since 1991 the data collection process has been administered by the National Opinion Research Center at the University of Chicago.} The objective of the survey is to characterize the financial situations of a set of households referred to as the Primary Economic Units (PEUs), where ``the PEU consists of an
economically dominant single individual or couple (married or living
as partners) in a household and all other individuals in the household
who are financially interdependent with that individual or couple.''\footnote{See the SCF codebook at \href{https://www.federalreserve.gov/econres/files/codebk2016.txt}{https://www.federalreserve.gov/econres/files/codebk2016.txt}  (accessed on April 15, 2019).}  

The SCF was initiated in 1982 and over the years
has become one of the primary data sources in studying consumer finances. It provides exhaustive categorization and detailed information of a variety of household financial products.\footnote{The SCF collects information on 
checking, brokerage, savings, and money market accounts; certificates of deposit; savings bonds and other types of bonds; mutual funds; publicly-traded
stocks; annuities, trusts, and managed investment accounts; IRAs and
Keogh accounts; life
insurance policies; and other types of financial and non-financial assets; credit card debt;
vehicle loans and other types of consumer loans; mortgages, lines of credit, and other loans.} While the most comprehensive data are collected on household portfolios, the survey also provides supplementary information on a wide range of demographic and socio-economic 
characteristics such as sex, age, race, ethnicity, family size, homeownership status, and employment history. 

Since the survey oversamples the upper tail of wealth distribution, it is also one of the primary data sources used in studying economic inequality. However, as emphasized by the Board, ``even under ideal operational conditions, the
measurements of the survey are limited in a fundamental way by the
fact that it is based on a sample of respondents rather than the
entire population.''\footnote{See the SCF codebook at \href{https://www.federalreserve.gov/econres/files/codebk2016.txt}{https://www.federalreserve.gov/econres/files/codebk2016.txt}  (accessed on April 15, 2019).} 

In this paper, I use the SCF data between 1988 and 2018, where the 2018 SCF is the latest available data set at the time of writing. The data sets from 1982 and 1985 are not included as they do not provide enough information to reliably estimate the main sources of variation in the SCF point estimates.

\subsection{Estimation of the TSE}
\label{subsec:SCF_estimation1}
With respect to Section \ref{sec:tse}, SCF data users with access to publicly available data files and supplementary materials can estimate two types of error, sampling and nonresponse. Quantifying other types of error such as specification, processing, frame, and measurement would constitute a nontrivial task that, in most instances, would require access to undisclosed information regarding specifics of the data editing process or construction of the sample frame, and as such is beyond the scope of the present paper.
\subsubsection{Sampling error}
\label{subsubsec:sampling_error}
In order to protect respondents' confidentiality, specifics regarding the SCF sampling design  are not disclosed to the general public. This  disclosure avoidance procedure has the objective of minimizing the risk of a third party revealing the identity of a survey respondent based on the sampling-specific information such as selection probability, sampling strata, and primary and secondary sampling units.

An important implication of this disclosure avoidance procedure is that sampling error cannot be estimated using standard statistical techniques or built-in functions in software such as STATA, SAS, SUDAN, or AM Statistical Software. Instead, I estimate the SCF sampling error using bootstrapped sample replicates, generated by the Board for all survey years between 1988 and 2018. The replicates are constructed based on the actual SCF sampling design (see Section A.1 in Appendix A of the Online Supplementary Material) and  are provided to the general public in the form of materials supplementary to the main data set. The main reason for providing these replicates is to facilitate the estimation of sampling error by all SCF data users, who lack access to the undisclosed and highly confidential information regarding the specifics of the SCF sampling design.

Let $\theta$ denote an unknown population parameter, and let $\hat{\theta}$ denote the estimate of $\theta$ computed in the main data set. Moreover, let $\hat{\theta}_l$ denote  the estimate of $\theta$ obtained in the $l$th bootstrapped sample replicate (as opposed to the main data set), where $l:1\rightarrow L$, and $L=999$.\footnote{The reason for choosing 999 bootstrapped sample replicates instead of 1,000 is that, as indicated in \citet{hall_1986_on_the_number}, with 1,000 repetitions (unlike 999), coverage probability of 90 percent bootstrap confidence interval is biased by approximately 0.001.} It follows that a sampling error of $\hat{\theta}$ is given by a sample standard deviation of $\left\lbrace \hat{\theta}_l\right\rbrace_{l=1}^L$,
\begin{align}
\hat{\sigma}_{1,\hat{\theta}}=\sqrt{\frac{1}{L-1}\sum_{l=1}^L\left(\hat{\theta}_{l}-\frac{1}{L}\sum_{l'=1}^L\hat{\theta}_{l'}\right)^2}.
\label{eq:sampling_error}
\end{align}

For illustration, consider a problem of estimating the sampling error of the estimate of the top 10 percent income share. The estimation consists of two steps. In the first step, I estimate $\theta$ in each bootstrapped sample replicate, which yields a total of $L$ replicate-dependent estimates $\hat{\theta}_l$. In the second step, I estimate the sampling error of $\hat{\theta}$ by computing the sample standard deviation of $\left\lbrace \hat{\theta}_l\right\rbrace_{l=1}^L$.

\subsubsection{Nonresponse error}
\label{subsubsec:nonresponse_error}
In addition to sampling error, SCF data users can estimate two types of nonresponse: item nonresponse and incomplete response.\footnote{Unit nonresponse is compensated for with nonresponse-adjusted sampling weights computed by the Board based on undisclosed selection probabilities, specifics regarding the sample frame, and population totals estimated from the Current Population Survey \citep[for more details see][]{kennickell_1997_consistent}. }  Across all survey years between 1988 and 2018, the SCF contains $M=5$ multiple imputations for virtually all variables (dichotomous and continuous) initially coded as either partially or completely missing.\footnote{Note that since the $M$ imputations are stored as five successive records for each survey respondent, the number of observations in each data set is mechanically inflated by a factor of five.}$^,$\footnote{For a general discussion on multiple imputation for
nonresponse in surveys see \citet{rubin_1987_miltiple}. For more information on multiple imputation in the SCF see \citet{kennickell_1998_multiple}.}$^,$\footnote{See Section A.2 in Appendix A of the Online Supplementary Material for more details on partially missing values in the SCF.}  This data feature allows users to account for the uncertainty associated with completely and partially 
missing data by estimating the variability between the multiply imputed data sets as
\begin{align}
\hat{\sigma}_{2,\hat{\theta}}=\sqrt{\frac{1}{M-1}\sum\limits_{m=1}^M\left(\hat{\theta}_{m}-\hat{\theta}\right)^2},
\label{eq:imputation_error}
\end{align}
where $\hat{\theta}_m$, $m:1\rightarrow M$ denotes the estimate of $\theta$ in the $m$th imputation.
\subsubsection{Standard error}
\label{subsubsec:standard_error}
After estimating sampling and imputation errors, I estimate the standard error of $\hat{\theta}$ using Rubin's  estimator, given by:
\begin{align}
\hat{\sigma}_{\hat{\theta}} = \sqrt{\hat{\sigma}^2_{1,\hat{\theta}}+\hat{\sigma}^2_{2,\hat{\theta}}\left(1+M^{-1}\right)}.
\label{eq:standard_error}
\end{align}
More details regarding Rubin's variance estimator can be found in Section A.3 in Appendix A of the Online Supplementary Material.

Since the above estimator of the standard error of $\hat{\theta}$ (see equation \ref{eq:standard_error}) accounts for only two types of error, sampling and nonresponse, it can be thought of as a lower bound on the unknown standard error of $\hat{\theta}$, say $\sigma_{\hat{\theta}}$. Importantly, the degree to which $\hat{\sigma}_{\hat{\theta}}$ underestimates $\sigma_{\hat{\theta}}$ depends on the magnitudes of the four nonsampling  errors that my estimation procedure does not account for. Given the high level of expertise of the Board in designing and supervising the survey, I assume three of these errors---specification, processing, and frame---to be fairly marginal. The only other type of error that may cause $\hat{\sigma}_{\hat{\theta}}$  to severely underestimate $\sigma_{\hat{\theta}}$  is measurement error, and more specifically, respondent-related measurement error, which I thoroughly discuss in Section A.4 in Appendix A of the Online Supplementary Material.

\section{Uncertainties in administrative data on tax returns}
\label{sec:income_inequality_in_administrative_data}
Starting from the early 1960s, the SOI Division of the IRS began to draw an annual sample of the Individual and Sole Proprietorship (INSOLE) tax returns. The INSOLE sample contains detailed information on taxpayers' incomes, deductions, exemptions, taxes, and credits, and thereby constitutes a micro-level database for tax policy purposes. In its present form, the INSOLE sample contains highly sensitive information that, if made publicly available, could risk the exposure of taxpayers' identities. Therefore, in order to ensure full confidentiality of the entire sample, access to the INSOLE is highly restricted and only granted to a handful of agencies, such as the Treasury Department or Congress. 

Since access to the INSOLE is strictly limited, the SOI Division annually creates another sample of tax returns commonly referred to as the PUF. The PUF is annually sub-sampled from the INSOLE and subjected to a number of disclosure avoidance procedures such as blurring, rounding, deleting, and modifying. These techniques have the objective of ensuring that no taxpayer can be identified from the PUF upon its release to the general public.\footnote{See Sections B.1 and B.2 in Appendix B of the Online Supplementary Material for a detailed description of the PUF and INSOLE sampling design.}  Hence, the PUF is accessible to much broader audiences, including academic and non-academic researchers, and constitutes one of the main data sets used in studies of inequality.

In the present paper, I rely upon the PUF data between 1991 and 2012, where  the 2012 PUF is the latest available data set at the time of writing.\footnote{Due to COVID-19, the release of the 2013 PUF is not known at the moment.}  As such, I analyze a twenty-two-year period that follows the last formal redesign of the INSOLE from the late 1980s and includes four years of the PUF after its latest revision, which took place in 2009.\footnote{``The revised design modifies which returns in the INSOLE sample are excluded from the PUF, changes the way the INSOLE sample is subsampled for the PUF, and aggregates all returns with a `large' value for any specified amount variable into a single record" \citep{bryant_2014_design}.}

\subsection{Estimation of the TSE}
\label{subsec:puf_total_survey_error}
Contrary to popular belief, the problem of data deficiencies pertains not only to self-reported survey data (such as the SCF) but also to data that comprise administrative records (such as the PUF). In fact, administrative data (which until very recently had been considered free of any source of error) are subject to the same types of error as survey data (whose accuracy is known to be
negatively affected by  sampling and nonsampling errors). Therefore, as \citet{groen_2012_sources} indicates, ``analyses of the quality of administrative data and reasons for differences between
administrative data and survey data are greatly needed.''\footnote{See Section \ref{sec:tse} for a general discussion on sampling and  nonsampling errors and Section B.3 in Appendix A of the Online Supplementary Material for a detailed description of different sources of error in the PUF.} 

In the present paper, I focus on the estimation of the PUF sampling error. As such, my analysis does not account for processing, nonresponse, measurement, specification, and frame errors. Whereas there is reason to assume that the first four types of error are marginal (see Section B.4 in Appendix  B of the Online Supplementary Material for  qualitative evidence supporting this claim), frame error may not be inconsequential. 

In order to illustrate why the PUF frame error may matter, consider work by \citet{income_2003_piketty}, where the authors impute income of non-filers as a fixed fraction of filers' average income for all years between 1946 and 1998. Even though this particular imputation procedure has the objective of matching the ratio (of 75--80 percent) of gross income reported on tax returns and total personal income estimated in national accounts, it remains highly arbitrary. As such, this and other imputation and/or estimation procedures aimed at ``filling the frame'' with  information on non-filers introduce additional and non-negligible sources of error into the analysis.
%
%Moreover, when estimating  wealth shares using the PUF, I account for an additional source of error which I refer to as ``modeling error.'' This type of error occurs in the PUF  as a result of relying upon a capitalization model in order to infer wealth concentration measures from data on taxpayers' income.

\subsubsection{Sampling error} Since the IRS does not provide data users with bootstrapped sample replicates, in order to estimate the PUF sampling error I first generate $L=999$ bootstrapped sample replicates based on the publicly available information on taxpayers' strata and stratum-specific probability of selection.  For brevity, let $\mathcal{S}$ denote the PUF sample of taxpayers, and assume that $\mathcal{S}$ comprises $J$ mutually exclusive and collectively exhaustive strata such that 
\begin{align}
\mathcal{S}=\cup_{j=1}^J\mathcal{S}_j,
\label{eq:sample_set_definition}
\end{align}
where $\mathcal{S}_j\cap\mathcal{S}_{j'}=\emptyset$ for all $j\neq j'$.

Moreover, let $n$ denote the total sample size, and let $n_j$ be the number of taxpayers selected for the sample from stratum $j$. Since the strata are mutually exclusive and collectively exhaustive, it follows from equation \eqref{eq:sample_set_definition} that
\begin{align}
n=\sum_{j=1}^Jn_j.
\end{align}

Since across all tax years under consideration there exist strata with as low as 10 observations or fewer (see Table C.1 in Appendix C of the Online Supplementary Material), bootstrapping methods cannot be applied directly to $\left\lbrace \mathcal{S}_j\right\rbrace_{j=1}^J$. Instead, I first classify the $J$ strata into $J^{\star}\ll J$ clusters using the Partitioning Around Medoids (PAM) clustering procedure \citep[see][]{reynolds_1992_clustering}, where I determine the number of clusters in each tax year based on a silhouette  analysis. The clustering procedure uses as an input three stratification variables originally designated for the INSOLE sample: gross income, presence or absence of special forms and schedules, and the return's potential usefulness for tax policy modeling. Since the income variable is ordinal (successive income brackets) whereas the latter two are nominal, I use the Gower distance measure, which is applicable to a mix of ordinal and nominal variables. 

The clustering procedure results in a PUF sample of taxpayers $\mathcal{S}$ that comprises $J^*$ mutually exclusive and collectively exhaustive clusters such that
\begin{align}
\mathcal{S}=\cup_{j=1}^{J^{\star}}\mathcal{S}^{\star}_j,
\label{eq:clustering_set} 
\end{align}
where $\mathcal{S}_j^{\star}\cap\mathcal{S}_{j'}^{\star}=\emptyset$ for all $j\neq j'$.

Moreover, with $n^{\star}_j$ denoting the number of taxpayers in cluster $j$, it follows from equation \eqref{eq:clustering_set} that
\begin{align}
n=\sum_{j=1}^{J^{\star}}n^{\star}_j.
\end{align}

For example, in tax year 2008, I classify the 95 strata (with the minimum number of observations per stratum equal to 11) into 23 clusters (with the minimum number of observations per stratum equal to 184). Summary statistics for clustered strata in the remaining tax years (1991 through 2012) can be found in  Table C.1 in Appendix C of the Online Supplementary Material. 

After classifying taxpayers into $J^{\star}$ clusters, I draw $L=999$ independent bootstrapped sample replicates. Specifically, for each
sample replicate $l:1\to L$, I draw with replacement $n^{\star}_j$ sample observations from each cluster $j$,
such that the total number of observations in each sample replicate is equal to $n$.

Finally, in order to estimate the PUF sampling error, I follow the estimation procedure of the SCF sampling error outlined in Section \ref{subsubsec:sampling_error}. Let $\hat{\theta}$ denote the estimate of $\theta$ computed in the main data set, and let $\hat{\theta}_l$ denote  the estimate of $\theta$ obtained in the $l$th bootstrapped sample replicate (as opposed to the main data set). I estimate the sampling error of $\hat{\theta}$ by a sample standard deviation of $\left\lbrace \hat{\theta}_l\right\rbrace_{l=1}^L$ as
\begin{align}
\hat{\sigma}_{1,\hat{\theta}}=\sqrt{\frac{1}{L-1}\sum_{l=1}^L\left(\hat{\theta}_{l}-\frac{1}{L}\sum_{l'=1}^L\hat{\theta}_{l'}\right)^2}.
\label{eq:sampling_error}
\end{align}

\section{Income inequality measured in survey and administrative data}
\label{sec:income_inequality}
In the following, I first define the concept of  income and wealth used throughout the paper and next, describe the procedure for estimating top income and wealth shares in the SCF and the PUF. Note that for the PUF, the derivations that follow apply to every tax year between 1991 and 2012, and for the SCF, to all survey years between 1988 to 2018. The time
subscript $t$ is omitted for ease of notation.

\subsection{Income}
In this paper, I define income as gross income comprising all income items, except for
capital gains, prior to deductions. The reason for excluding capital gains 
is that ``realized capital gains are not an annual flow of income (in general, capital gains are realized by
individuals in a lumpy way only once in a while) and form a very volatile component of income with large
aggregate variations from year to year depending on stock price variations'' \citep{income_2003_piketty}. 

The aforementioned income measure  (as well as numerous alternative measures that could be applied without loss of generality, such as gross income \textit{including} capital gains) can be constructed using each of the two data sets under consideration, without the need to rely upon supplementary data and/or econometric modeling. Specifically, the SCF collects information on households' income during the SCF interview process (\textit{In total, what was your annual income from
dividends, before deductions for taxes and
anything else?}), whereas the PUF compiles income data from a sample of filed tax returns (Form 1040).\footnote{For the SCF, I compute gross income less capital gains as a sum of salaries and wages before deductions for taxes (variable X5702); income from a sole proprietorship or a farm (X5704); income from other businesses or investments; net rent, trusts, or
royalties (X5714); income from non-taxable investments (X5706); income from other interest (X5708); income from dividends (X5710); income from Social Security or other pensions; annuities, or other disability or retirement programs (X5722); income from unemployment or worker's compensation (X5716); income from child support or alimony (X5718);  and income from other sources (X5724).}$^,$\footnote{For the PUF, I compute gross income less capital gains as a sum of wages and salaries (line 7 on Form 1040); taxable interest
(line 8a); tax-exempt interest (line 8b); ordinary dividends (line 9a); taxable refunds, credits, and offsets of state and local income taxes (line 10); alimony received (line 11); business and farm income (lines 12 and
18); IRA distributions, pensions and annuities, unemployment compensation, and social security benefits (lines 15b, 16a, 19, and 20a); and rental real estate, royalties, partnerships, S corporations, and trusts (line
17).} The resulting operational definitions of income are virtually identical for the two data sets, differing only with respect to two income components: the SCF reports both taxable and nontaxable IRA distributions and all other sources of income, the PUF reports only taxable amounts from line 15a on Form 1040 and does not provide information on all other sources of income from line 21.\footnote{Note that for comparability with the PUF, I do not include welfare assistance in SCF measure of wealth.} 
\subsection{Wealth}
\label{puf_wealth}
Throughout my analysis, I define wealth as total assets less total debt. Since the SCF survey participants are asked detailed questions about their asset and liability holdings,  measuring wealth in the SCF is straightforward and boils down to a simple accounting exercise.\footnote{See Figure C.1 in Appendix C of the Online Supplementary Material for a detailed description of the construction of the SCF wealth measure.} In contrast, the PUF---a sample of individual income tax returns---provides limited information on taxpayers' wealth, which results from the fact that many asset and liability holdings are not reported on a tax form. For example, since for many taxpayers standard deductions are more effective at reducing financial burden than are itemized deductions, only a small fraction of homeowners deduct mortgage interests on their tax returns. 

In order to construct comprehensive wealth measures using the PUF (as well as the INSOLE or any other tax-level data) it is necessary to rely upon auxiliary data sources and numerous modeling assumptions. In this paper, I utilize a capitalization model from \citet{saez_2016_wealthinequality} and later re-visit in \citet{bricker_2018_how_much}, where  taxpayers' wealth is estimated by ``capitalizing''  asset income with an asset-specific rate of return. Specifically, as summarized in \citet{saez_2016_wealthinequality}, for each asset class I estimate a capitalization factor that maps the total flow of tax income to the amount of wealth from the household balance sheet of the Financial Accounts of the United States. Then, I estimate wealth of each tax payer by multiplying their reported incomes by the corresponding capitalization factors.

Let the wealth of taxpayer $i$ be defined as
\begin{align}
\widehat{wealth}_i = \widehat{nonfin}_i + \sum_{a=1}^A\frac{income_{i,a}}{\hat{r}_a},
\label{eq:wealth_pred_sz}
\end{align}
where $income_{i,a}$ denotes the income of taxpayer $i$ generated by asset $a=1,\cdots, A$, $\hat{r}_a$ denotes the estimated rate of return on asset $a$, and $\widehat{nonfin}_i$ is the estimate of taxpayer $i^'$s nonfinancial wealth.

The rate of return on asset $a$ is estimated by computing a ratio of 
the household stock of asset $a$ reported in the Financial Accounts, say $FA_a$, to the $a'$s realized capital income measured in the PUF,
\begin{align}
\hat{r}_a=\frac{\sum_{i=1}^n income_{i,a}}{FA_a}.
\label{eq:ror}
\end{align}

Following \citet{saez_2016_wealthinequality}, I organize assets from the Financial Accounts into seven categories: (1) taxable interest-bearing assets, (2) non-taxable interest-bearing assets, (3) dividend-generating assets, (4) assets generating profits of S corporations, (5) assets generating royalty income and profits of partnerships and C corporations, (6)  tenant-occupied real estate assets less mortgages, and (7) privately held and employer-sponsored pension assets.\footnote{In order to construct seven aggregate asset categories, it is often necessary to combine numerous lines from multiple  tables reported in the Financial Accounts.}$^,$\footnote{Details regarding the construction of the remaining asset classes can be found on my personal website as well as in the Online Appendix to \citet{saez_2016_wealthinequality}.} 

For illustration, consider the following problem of determining assets of taxpayer $i^{\star}$ with reported income from dividends of \$6,710. First, I estimate rate of return on dividend-generating assets, say $\hat{r}_{div}$, using equation \eqref{eq:ror}, with the denominator $FA_{div}$ computed as a sum of directly held equities (FL153064105), equities indirectly held through mutual funds (FL653064155), and the share of equities in money market funds (estimated based on FL153034005, FL634090005, and FL633062000), less equities held by nonprofit organizations and mutual funds held in IRAs. Then, I estimate dividend-generating assets of taxpayer $i^{\star}$ by dividing her/his dividend income of \$6,710 by the estimated rate of return. In 2012, the estimated rate of return was equal to 0.03411, implying a total of \$196,717 in dividend-generating assets for taxpayer $i^{\star}$.\footnote{Since the Financial Accounts are revised on a quarterly basis, I use tables from the first quarter of 2020, which contain the latest data available at the time of writing.}

In this paper, I consider three sets of PUF estimates, one generated under a homogeneity assumption imposed on all rates of return under consideration, as in equation \eqref{eq:ror}, and two sets of estimates constructed under a heterogeneity assumption, where I assume homogeneous rates of return on all income-generating assets except
for those that generate taxable interests. As emphasized by \citet{bricker_2018_how_much} ``implied rate of return on taxable interest-bearing assets in \citet{saez_2016_wealthinequality} is much lower than market rates from the 10-year Treasury yield or Moody's  Aaa corporate bond---the type of taxable interest-bearing assets that are held by wealthy families \citep{bricker_2016_measuring, kopczuk_2015_what_do_we_know}.'' Therefore, following \citet{bricker_2018_how_much}, I consider a scenario, where I assign a higher rate of return, say $\hat{r}_{a,\mathcal{A}}$, to the top 1 percent of the wealth distribution, and a lower rate, say $\hat{r}_{a,\mathcal{B}}$, to the bottom 99 percent, such that
\begin{align}
FA_a = \frac{\sum_{i\in\mathcal{A}} income_{i,a}}{\hat{r}_{a,\mathcal{A}}} + \frac{\sum_{i\in\mathcal{B}} income_{i,a}}{\hat{r}_{a,\mathcal{B}}},  
\end{align}
where $\mathcal{A}$ denotes a set comprised of the top 1 percent of the wealth distribution and $\mathcal{B}$ a set comprised of the bottom 99 percent.

%As such, this paper replicates and builds upon  \citet{bricker_2018_how_much} in order to demonstrate the significance of modeling errors in studies on top wealth inequality.
Note that the resulting operational definitions of wealth in the SCF and the PUF differ  with respect to three asset categories: defined
pension plans and term life insurance policies, which are included in the PUF but not in the SCF, and durable goods (e.g., vehicles), which are included in the SCF but not in the PUF.\footnote{\citet{sabelhaus_2019_are_disappearing_employer_pensions} estimate defined benefit plans for the SCF survey participants using an actuarial
present-value model. The model utilizes information on benefit amounts for those SCF respondents' currently collecting a pension, the expected timing and amount of future pension benefits from a past job for those who are entitled to a benefit but are not yet collecting it, and the current wage,
age, sector (private or public), and the number of years in the plan of those who have a plan
tied to their current job but are not yet receiving benefits.}

\subsection{Estimation}
In this section, I describe the estimation procedure for top income and wealth shares using the two data sets under consideration, the SCF survey data and the PUF sample of administrative tax records.

Let $g_i$ denote either income or wealth of observation $i:1\rightarrow n$ (in either the SCF or the PUF), and  let $w_i$ denote sampling weight of $i$ such that
\begin{align}
\sum_{i=1}^nw_i=N,
\end{align}
where $N$ denotes the number of units in the underlying population of interest. For the SCF, $N$ is equal to the total number of PEUs. For the PUF,  $N$ is equal to the total number of taxpayers.

Let $g_{\left( j\right)}$ be the $j^{\text{th}}$-order statistic of $g_i$, with $w_j$ denoting the sampling weight associated with $g_{\left( j\right)}$.\footnote{Order statistics of $g_i$ are ranked in ascending order of magnitude \citep[see, e.g.,][]{david_2003_order} with $g_{\left(1\right)}=\min_j\left\lbrace g_i\right\rbrace$ and $g_{\left(n\right)}=\max_j\left\lbrace g_i\right\rbrace$.} Moreover, let $m_k$ denote the number of observations in the bottom $100k$ percent defined as
\begin{align}
m_k=kN,
\label{eq:mk_definition}
\end{align}
where $k\in \mathcal{K}$ and $\mathcal{K}=\left\lbrace 0.9, 0.95, 0.99 , 0.995, 0.999, 0.9999\right\rbrace$.

For example, consider tax year 2008, with the number of taxpayers equal to N = 142,580,866.
It follows from equation \eqref{eq:mk_definition} that $m_{0.9}=12,322,809$ and $m_{0.9999}=141,155,090$.

Next, let  $r_k$ denote the unknown (income or wealth) share of the \textit{bottom} $100k$ percent, and let $p_k$ be the unknown share of the \textit{top} $100\left(1-k\right)$ percent defined as
\begin{align}
p_{k} =1 - r_k.
\label{eq:pk_definition_as1_-r_k}
\end{align}

The estimation procedure consists of two main steps. In the first step, I determine an index value $j_k^{\star}\in\left\lbrace 1,\cdots, n-1\right\rbrace$ such that
\begin{align}
\sum_{j=1}^{j_k^{\star}}w_j \leq m_k \leq \sum_{j=1}^{j_k^{\star}+1}w_j,
\end{align}

which allows me to estimate the lower bound on $r_k$  as
\begin{align}
\underline{r}_k=\frac{\sum_{j=1}^{j_k^{\star}}w_{j}g_{\left(j\right)}}{\sum_{j=1}^{n}w_{j}g_{\left(j\right)}},
\end{align}
and the upper bound on $r_k$ as
\begin{align}
\overline{r}_k=\frac{\sum_{j=1}^{j_k^{\star}+1}w_{j}g_{\left(j\right)}}{\sum_{j=1}^{n}w_{j}g_{\left(j\right)}}.
\end{align}

In the second step, I estimate $r_k $ using linear interpolation between $\underline{r}_k$ and $\overline{r}_k$:
\begin{align}
\hat{r}_k=\underline{r}_k+\omega_k\left(\overline{r}_k-\underline{r}_k\right),
\end{align}
where
\begin{align}
\omega_k=\frac{m_k-\sum_{j=1}^{j_k^{\star}}w_j}{w_{j_k^{\star}+1}}.
\end{align}

It follows from equation \eqref{eq:pk_definition_as1_-r_k} that the estimator of $p_k$ is given by
\begin{align}
\hat{p}_k=1-\hat{r}_k.
\end{align}

In relation to Section \ref{subsubsec:nonresponse_error}, it is important to note that since the SCF data are multiply imputed for missing values, the aforementioned estimation procedure needs to be repeated  separately for each of the $M$ imputed SCF data sets. This results in $M$ estimates of $p_k$, which I denote as $\hat{p}_{k,m}$, $m:1\rightarrow M$. 
The grand estimate of $p_k$ is then obtained by averaging over $\left\lbrace \hat{p}_{k,m}\right\rbrace_{m=1}^M$ such that:
\begin{align}
\hat{p}_k=\frac{1}{M}\sum_{m=1}^M\hat{p}_{k,m}.
\end{align}
%%%%%%%%%%%%%%%%%%%%%%%%%%%%%%%%%%%%%%%%%%%%%%%%%%%%%%%%%%%%%%%%
\section{Empirical results on income shares}
\label{sec:empirical_results_income}
In the following, I discuss my empirical results regarding the estimation of the income shares within the top 10 percent. In Section \ref{subsec:number_of_observations}, I compare the number of observations above top income fractiles estimated using the the SCF and the PUF. In Sections \ref{subsec:income_point_estimates}--\ref{subsec:income_inequality_great_recession}, I focus on point estimates, standard errors, and the long- and short-term dynamics in income inequality. In Section \ref{subsec:cross-correlations}, I establish a statistical link between the SCF and the PUF point estimates for top income shares. Finally, in Section \ref{subsec:income_Key findings}, I summarize my key findings and conclude.
\subsection{Number of observations}
\label{subsec:number_of_observations}
My first set of results pertains to the number of observations in the SCF and the PUF. Across all of the years under study, I find large differences in the  number of observations between the SCF and the PUF. For example, as indicated in Table \ref{tab:weighted_number_of_obsrevations}, the number of observations in the SCF in 2012 accounts for just 3.5 percent of the total number of observations available in the PUF. Moreover, since the number of observations increases proportionately across the two data sets, the ratio of the number of the SCF sample observations to the number of the PUF sample observations is fairly stable across years with a minimum of 3 and a maximum of 4.5 percent.\footnote{See Table C.2 in Appendix C of the Online Supplementary Material for the number of observations in the SCF and the PUF.} 
% Table generated by Excel2LaTeX from sheet 'Number of obs inc comparison'
\begin{table}[htbp]
  \centering
  \caption{Ratio of the SCF number of observations to the PUF number of observations for income}
     \resizebox{.8\textwidth}{!}{\begin{tabular}{C{1.5cm}C{3.0cm}C{1.5cm}C{1.5cm}C{1.5cm}C{1.5cm}C{1.5cm}C{1.5cm}} \toprule
    \multicolumn{1}{c}{\multirow{2}[0]{*}{\textbf{Year}}} & \multicolumn{1}{c}{\multirow{2}[0]{3.0cm}{\textbf{Unweighted: total sample}}} & \multicolumn{6}{c}{\textbf{Weighted: above the \boldmath$k$ income fractile}} \\
    \cline{3-8}
          &       & \textbf{90} & \textbf{95} & \textbf{99} & \textbf{99.5} & \textbf{99.9} & \textbf{99.99} \\ \midrule
    1991  & 3.4   & 2.0   & 1.8   & 1.6   & 1.4   & 1.3   & 1.2 \\
    1994  & 4.5   & 2.8   & 2.5   & 2.0   & 1.8   & 1.4   & 1.3 \\
    1997  & 3.9   & 2.3   & 2.0   & 1.4   & 1.3   & 1.0   & 1.6 \\
    2000  & 3.0    & 1.6   & 1.4   & 1.0   & 0.9   & 0.8   & 1.5 \\
    2003  & 3.4   & 2.0   & 1.8   & 1.5   & 1.5   & 1.4   & 2.3 \\
    2006  & 3.0    & 1.9   & 1.7   & 1.6   & 1.6   & 1.9   & 7.2 \\
    2009  & 4.2   & 2.7   & 2.4   & 1.7   & 1.6   & 1.8   & 7.2 \\
    2012  & 3.5   & 2.1   & 1.7   & 1.3   & 1.3   & 2.1   & 5.8 \\\bottomrule
    \end{tabular}}%
  \label{tab:weighted_number_of_obsrevations}%
  \\\begin{tabular}{@{}c@{}} 
\multicolumn{1}{p{\columnwidth -1in}}{\footnotesize  Ratios are expressed in percent}
\end{tabular}
\end{table}%

In addition,  I observe large differences not only in the total number of observations but also in the weighted number of observations, including the far  right tail of income distribution. For example, the weighted number of observations above the 90 income fractile in the SCF in 2012 accounts for only 2.1 percent of the weighted number of observations above the 90 income fractile in the PUF.  More generally, between 1991 and 2012, this ratio varied from a minimum of 1.6 percent to a maximum of 3 percent. 

The number of observations in the SCF is small not only in relative terms when compared to the abundance of data in the PUF but also in absolute terms. In most of the years under study, I observe the weighted number of observations in the SCF above the 99.9 and 99.99 income fractiles to be less than three and one hundred, respectively, and hence, to be insufficient for a reliable estimation of the income shares of the top 0.1 and 0.01 percent.\footnote{See Table C.3 in Appendix C of the Online Supplementary Material for the weighted number of observations in the far right tail of the income distribution.} This result shows that the problem of a small number of observations in the SCF escalates when the focus of the analysis shifts from the mean or median to top income fractiles.

%All in all, I find the number of observations in the SCF above the 99.5 income fractile insufficient for a reliable estimation of the income shares of the top 0.1 and 0.01 percent. 

%%%%%%%%%%%%%%%%%%%%%%%%%%%%%%%%%%%%%%%%%%%%%%%%%%%%%%%%%%%%%%%%%%%%%%
%new
\subsection{Point estimates} 
\label{subsec:income_point_estimates}
Having compared the number of observations in the SCF and the PUF, in the following, I focus on point estimates. 
\begin{figure}[htb!]
\centering
\includegraphics[scale=0.55]{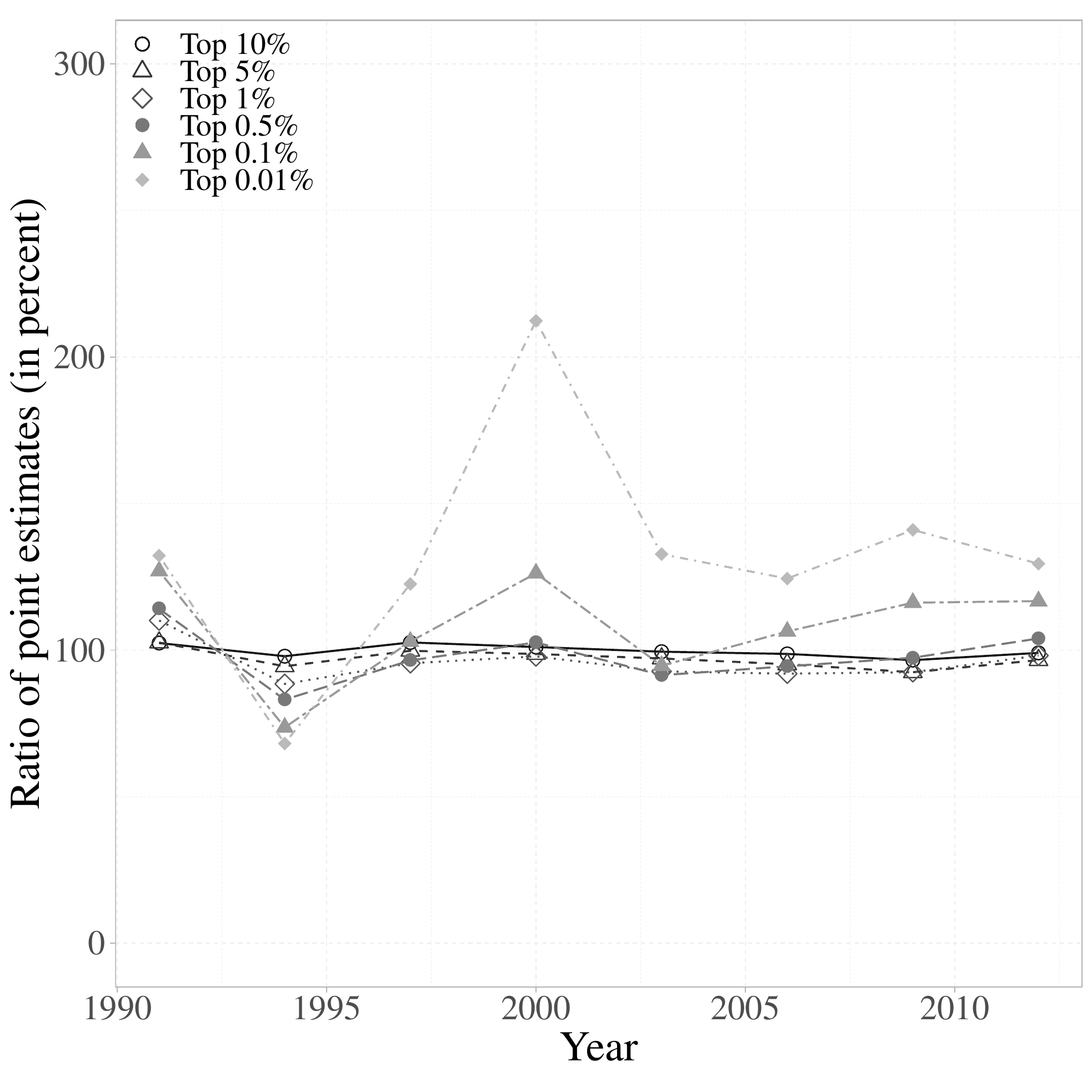}
\caption{Ratio of the PUF point estimate to the SCF point estimate for the income shares within the top 10 percent}
\label{fig:income_ratio_of_point_estimates}
\end{figure}
In particular, I compute ratios of the
PUF to the SCF point estimates for the six income shares under consideration percent between 1991 and 2012. My analysis suggests that the SCF and the PUF point estimates concur with regard to less granular income shares (such as the top 10, 5, 1, and 0.5 percent), with the SCF point estimates only marginally above those obtained using the PUF. However, with respect to the more granular income shares of the top 0.1 and 0.01 percent, the two data sets greatly disagree. Specifically, ratios of point estimates vary from a minimum of 70 to a maximum of 210 percent, with the SCF point estimates being systematically below those obtained using the PUF. Whether this is driven by  the small number of observations in the SCF (above the 99.9 and 99.99 income fractiles) or is related to a potential under-reporting of incomes by the SCF respondents from the top 0.1 and 0.01 percent of income distribution is beyond the scope of the current study. I will address this issue in a follow-up research project centered on measurement error arising from under- and over-reporting in financial surveys.

\subsection{Relative magnitudes of sampling error across data sets}
\label{subsec:Income_analysis_Relative magnitudes of sampling error across data sets}
In the present section, I analyze the ratios of the Coefficients of Variation (CVs) in the PUF to those in the SCF for the six income shares of the top 10 to the top 0.01 percent between 1991 and 2012.\footnote{CV is defined as sampling error standardized for point estimate.} 
\begin{figure}[hbt!]
\centering
\includegraphics[scale=0.55]{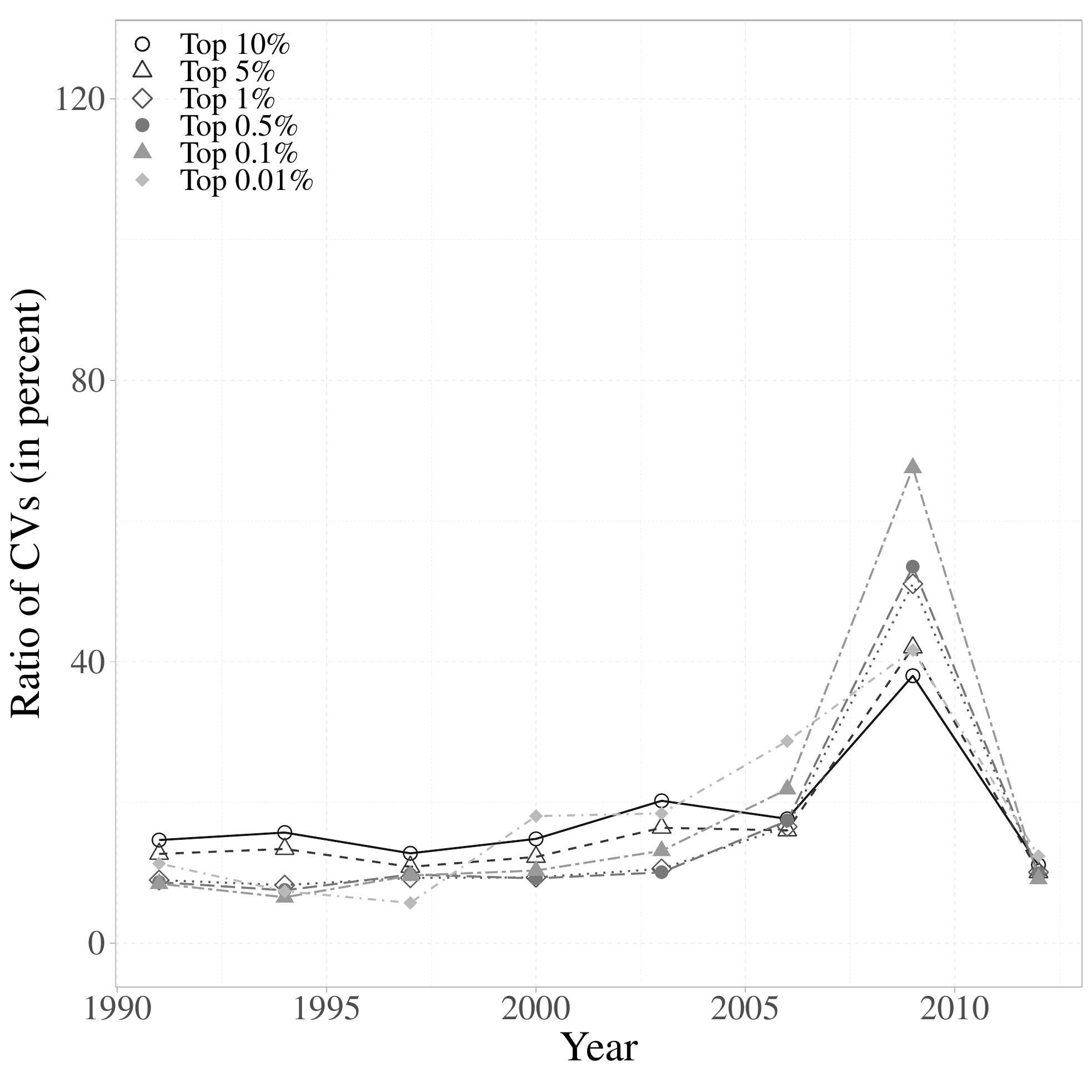}
\caption{Ratio of the PUF coefficient of variation to the SCF coefficient of variation for the income shares within the top 10 percent}
\label{fig:income_ratios_of_CV}
\end{figure}
My results, illustrated in Figure \ref{fig:income_ratios_of_CV},  indicate that the SCF sampling errors are much larger than those in the PUF for all years and income shares under consideration. Except for 2009, when I observe an unprecedented one-time increase in the PUF sampling error, CVs in the SCF are at least five times larger than those in the PUF. Note that the observed unprecedented one-time increase in the PUF sampling error in 2009 may have resulted from the change to the PUF sub-sampling design that occurred in 2009 or be related to the impact of the 2007--2009 Great Recession on households' financial situations. To conclude, I find the observed magnitudes of discrepancy in the sampling errors of the SCF and PUF a decisive factor in choosing between the two data sets for conducting studies of top income inequality.

\subsection{Relative magnitudes of sampling error across income shares}
\label{subsec:income_Relative magnitudes of sampling error across income shares}
\begin{figure}[htb!]
\centering
\includegraphics[scale=0.5]{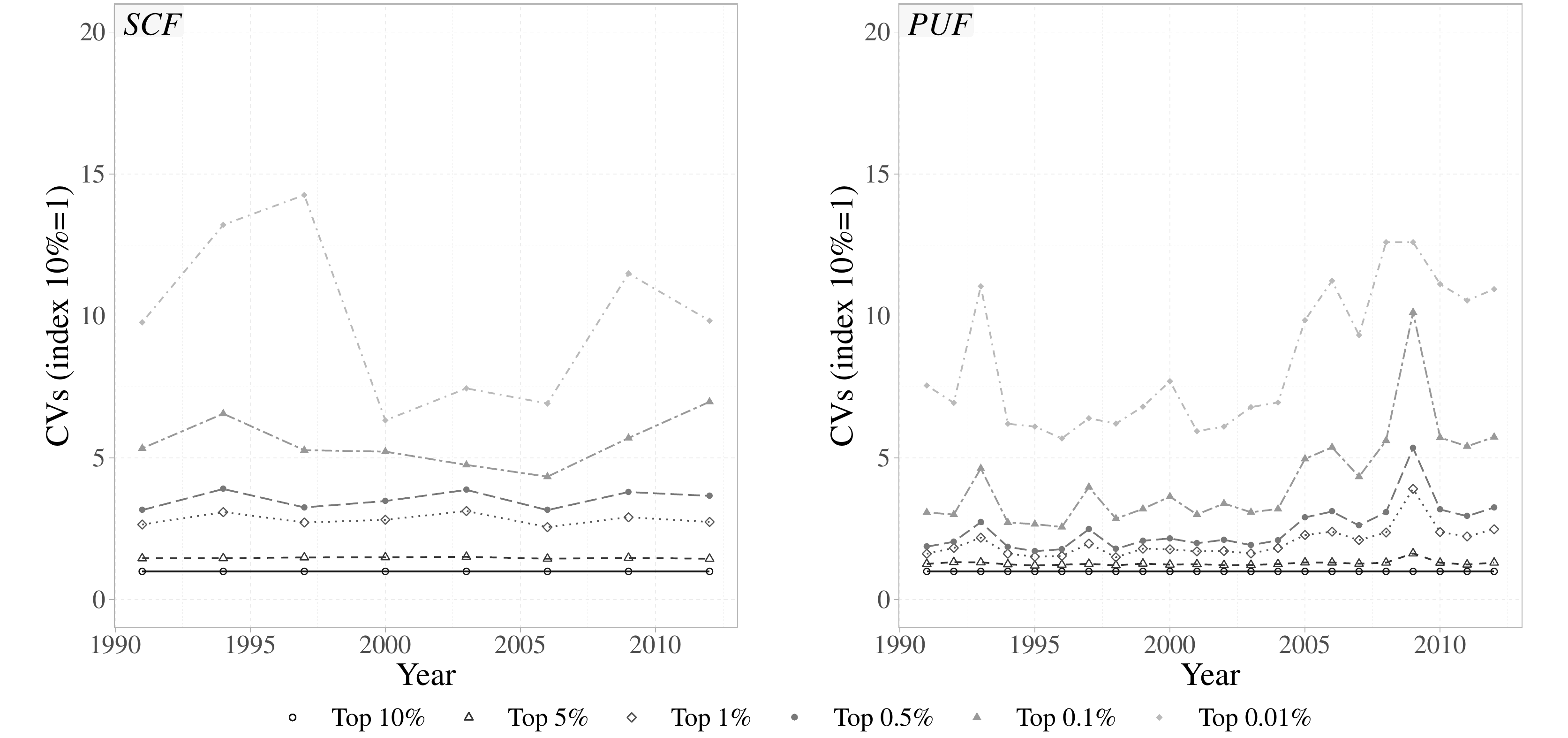}
\caption{CVs in the SCF and the PUF for the six income shares of the top 10 to the top 0.01 percent}
\label{fig:income_relative_CoeffVar_alt}
\end{figure}

In the following, to complement Section \ref{subsec:Income_analysis_Relative magnitudes of sampling error across data sets}, I compare the magnitudes of sampling errors \textit{within} a data set but \textit{across} estimated income shares. Specifically, for each of the two data sets under consideration, I compare the CVs for the income shares of the top 5, 1, 0.5, 0.1, and 0.01 percent to the CVs for the income share of the top 10 percent. As illustrated in Figure \ref{fig:income_relative_CoeffVar_alt}, an increase in CVs as the estimated income shares become more granular is not only inevitable but also substantial. However, this increase is much less pronounced in the PUF than it is in the SCF. Therefore, the PUF estimates for more granular income shares when compared to those for less granular income shares are estimated with substantially smaller error than those estimated using the SCF.

\subsection{Standard error decomposition}
\label{subsec:income_Standard error decomposition}
The present section breaks down SCF standard error for clearer analysis of leading sources of variation in the SCF estimates of top income inequality. In Table  \ref{tab:scf_error_decomposition}, I present ratios of sampling error to total standard error (that comprises both sampling and imputation errors) for the six income shares of the top 10 to the top 0.01 percent  between 1988 and 2018.  My analysis suggests that even though sampling error is the main source of variation in the SCF estimates of top inequality, imputation error is not to be discarded. Specifically, until the early 2000s, imputation error  accounted for at least 10 percent of total standard error, with the largest shares observed in the late 1980s and early 1990s.\footnote{Larger shares of imputation error in 1988 and 1991 can be explained by the fact that not until 1994 were the SCF respondents allowed the possibility of reporting partial (range) information on dollar amounts in an effort to reduce the number of completely missing cases. For more information on partially missing values in the SCF see Section A.2 in Appendix A of the Online Supplementary Material.} In the most recent years, the relative importance of imputation error has diminished, leaving sampling error as the sole source of variation in the SCF estimates of top income shares. Nevertheless, since imputation error was a significant contributor of the total variance  in earlier years, the SCF standard error that accounts for both sampling and imputation errors is likely to be on average more than five times larger than that constructed using the PUF. Note that this is the case since, even though this paper does not estimate the PUF imputation/nonreponse error, qualitative evidence suggests this source of error to be inconsequential for the PUF.
\begin{table}[hbt]
  \centering
  \caption{Ratio of sampling error to total standard error in the SCF estimates of the income shares within the top 10 percent}
    \resizebox{.7\textwidth}{!}{\begin{tabular}{C{1.7cm}C{1.7cm}C{1.7cm}C{1.7cm}C{1.7cm}C{1.7cm}C{1.7cm}} \toprule
    \multicolumn{1}{c}{\multirow{2}[0]{*}{\textbf{Year}}} & \multicolumn{6}{c}{\textbf{Income share of the top \boldmath$k$\%}} \\
        \arrayrulecolor{cadetgrey}\cline{2-7}
          & \textbf{10\%} & \textbf{5\%} & \textbf{1\%} & \textbf{0.5\%} & \textbf{0.1\%} & \textbf{0.01\%} \\\midrule
    1988  & 49.3  & 50.8  & 53.5  & 52.9  & 53.8  & 87.5 \\
    1991  & 82.8  & 77.0  & 88.4  & 91.8  & 89.3  & 92.6 \\
    1994  & 91.3  & 95.0  & 93.3  & 93.8  & 96.0  & 98.9 \\
    1997  & 97.1  & 95.4  & 89.7  & 81.8  & 77.0  & 91.1 \\
    2000  & 92.6  & 94.0  & 86.5  & 82.1  & 66.6  & 50.9 \\
    2003  & 78.9  & 81.7  & 88.8  & 92.4  & 92.7  & 97.2 \\
    2006  & 91.0  & 89.3  & 92.1  & 91.2  & 84.3  & 63.2 \\
    2009  & 95.7  & 93.5  & 97.7  & 98.3  & 96.3  & 97.3 \\
    2012  & 93.6  & 95.5  & 96.7  & 97.4  & 97.2  & 99.6 \\
    2015  & 97.2  & 98.4  & 99.2  & 99.6  & 98.5  & 99.5 \\
    2018  & 95.2  & 98.3  & 99.7  & 99.3  & 97.6  & 98.6\\\bottomrule
    \end{tabular}}%
  \label{tab:scf_error_decomposition}%
   \\\begin{tabular}{@{}c@{}} 
\multicolumn{1}{p{\columnwidth -2in}}{\footnotesize  Ratios are expressed in percent}
\end{tabular} 
\end{table}%

\subsection{Long-term trends in income inequality}
\label{subsec:income_Long-term trend in income inequality}
\begin{figure}[H]
\centering
\includegraphics[scale=0.85]{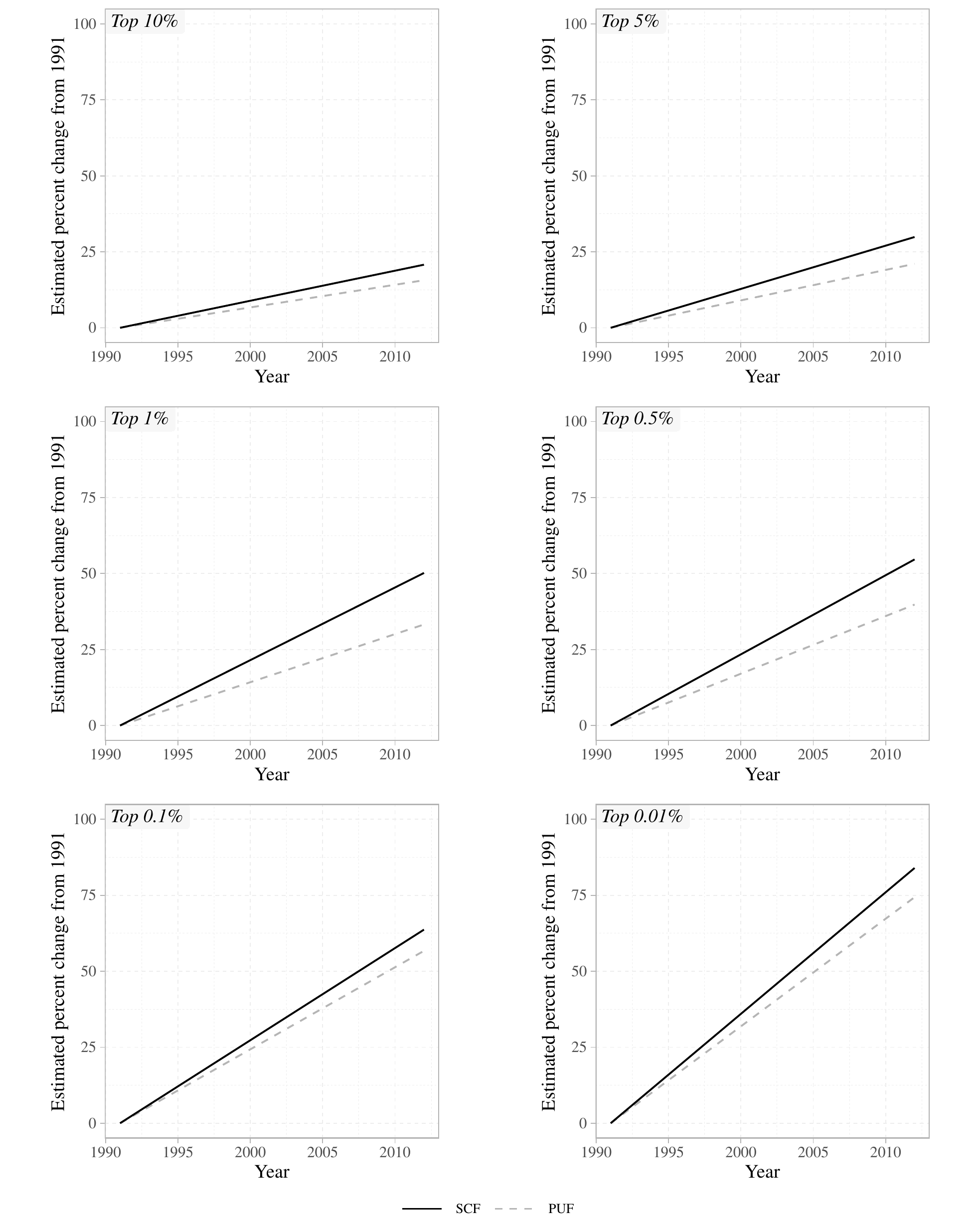}
\caption{Estimated increase in the income shares within the top 10 percent in the SCF and the PUF}
\label{fig:income_compact_trends_long}
\end{figure}

This section discusses long-term trends in income inequality. My two main objectives are to compare the estimated trend lines of the SCF and the PUF and to determine whether an observed increase in top income inequality between the early 1990s and early 2010s is statistically significant. In this exercise, I use data from 1991 through 2012 and regress estimated income shares (of the top 10 to the top 0.01 percent) on a constant and linear time trend using weighted least squares, with weights defined as reciprocals of squared standard errors.\footnote{Note that while the PUF standard error consists only of sampling error, the SCF standard error comprises both sampling and imputation errors.} As indicated in Figure \ref{fig:income_compact_trends_long}, I find that both the SCF and the PUF suggest a statistically significant increase in all six top-decile income shares under consideration (see Table C.5 in Appendix C of the Online Supplementary Material for the estimation details.). However, the two data sets do not fully agree with respect to the estimated increase in income shares, with the SCF trend lines being consistently steeper than those constructed using the PUF. Nevertheless, since the observed discrepancies are moderate and by no means extensive, the two data sets imply an increase of comparable magnitude in top income inequality between the early 1990s and early 2012.

\subsection{Income inequality and the Great Recession}
\label{subsec:income_inequality_great_recession}
\begin{figure}[htb!]
\centering
\includegraphics[scale=0.55]{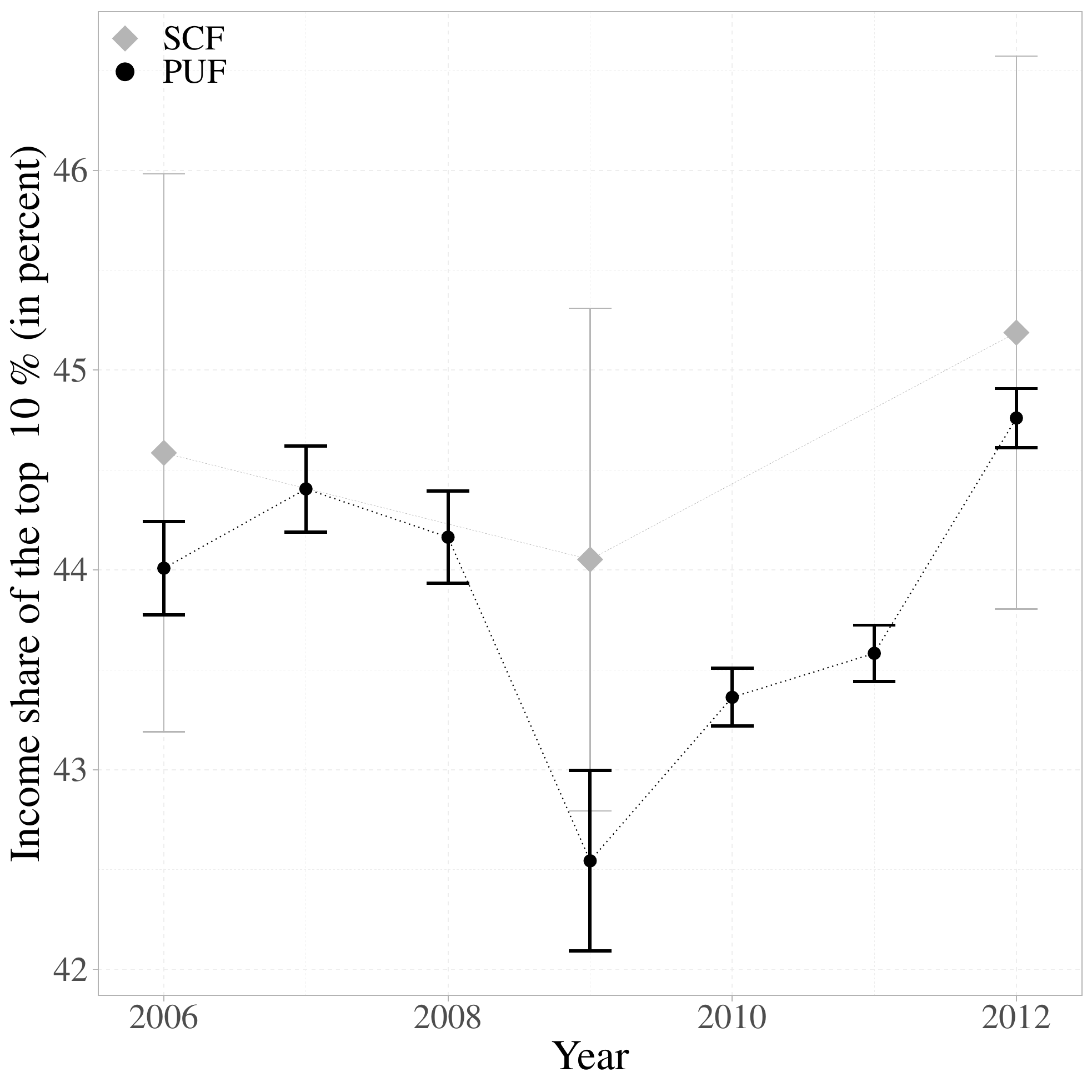}
\caption{Income shares of the top 10 percent before and after the 2007--2009 Great Recession in the SCF and the PUF. Error bars indicate 95 percent confidence intervals around point estimates}
\label{fig:income_great_recession}
\end{figure}

After discussing the long-term dynamics in income inequality, in the present section, I focus on a shorter time horizon before and after the 2007--2009 Great Recession. Specifically, in Figure \ref{fig:income_great_recession}, I present point estimates and their 95 percent confidence intervals for income shares of the top 10 percent, constructed 
using the SCF and the PUF between 2006 and 2012. I observe that the PUF estimates suggest a sharp and statistically significant decrease in incomes shares of the top 10 percent from 2007 to 2009, followed by a three-year long recovery to pre-recession levels. As noted by \citet{thompson_2018_top_income_concentration} ``the factors explaining the rise and fall in income concentration are not fully understood, but some of the most prominent explanations for rising top incomes highlight the role played by individuals---who may be `superstars' \citep{rosen_1981_economics_of_superstars} or `rent seekers' \citep{bivens_2013_pay_of_corporate_executives}---whose compensation is relatively volatile from one year to the next \citep[][among others]{bebchuk_2003_executive_compensation, kaplan_2013_it_is_the_market}.'' By contrast, the SCF does not support the claim of a statistically significant change in income shares of the top 10 percent in relation to the impact of the Great Recession.  Therefore, while the SCF can be used to analyze long-term dynamics and detect changes in income shares over longer-time horizons, it lacks statistical power to determine changes in income inequality over shorter-time horizons, including recessions and  economic expansions.

\subsection{Regression analysis}
\label{subsec:cross-correlations}
Lastly, I investigate whether there exists a statistical link between the SCF and the PUF. In this exercise, I regress the PUF point estimates on the SCF point estimates for each top-decile income share under consideration. As shown in Table \ref{tab:income_estimation_results_regression_PUF_on_SCF}, my analysis indicates strong correlation between the PUF and the SCF income shares of the top 10 to the top 0.5 percent.\footnote{For more granular income shares of the top 0.1 and 0.01 percent, the relationship in question is statistically insignificant.}  This result is of great importance since it opens up the possibility of merging the two data sets into one, which would result in a superior data set with plenty of observations (see the PUF) and a rich set of demographic and socio-economic characteristics (see the SCF). 
%new
% Table generated by Excel2LaTeX from sheet 'Regressions'
\begin{table}[htbp]
  \centering
  \caption{Estimation results from regressing the PUF on the SCF for the income shares within the top 10 percent}
    \resizebox{.9\textwidth}{!}{\begin{tabular}{lC{2cm}C{2cm}C{2cm}C{2cm}C{2cm}C{2cm}}\toprule
    \multirow{2}[0]{*}{\textbf{Regressor}} & \multicolumn{6}{c}{\textbf{PUF income share of the top \boldmath$k$\%}} \\
    \cline{2-7}
          & \textbf{10\%} & \textbf{5\%} & \textbf{1\%} & \textbf{0.5\%} & \textbf{0.1\%} & \textbf{0.01\%} \\\midrule
    \multirow{2}[0]{*}{Intercept} & 0.075$^{\textcolor{white}{\star\star\star}}$ & 0.061$^{\textcolor{white}{\star\star\star}}$ & 0.030$^{\textcolor{white}{\star\star\star}}$ & 0.021$^{\textcolor{white}{\star\star\star}}$ & 0.024$^{\textcolor{white}{\star\star\star}}$ & 0.017$^{\textcolor{white}{\star\star\star}}$ \\
          & \scriptsize (1.793)$^{\textcolor{white}{\star\star\star}}$ & \scriptsize (1.990)$^{\textcolor{white}{\star\star\star}}$ & \scriptsize (1.486)$^{\textcolor{white}{\star\star\star}}$ & \scriptsize (0.892)$^{\textcolor{white}{\star\star\star}}$ & \scriptsize (1.126)$^{\textcolor{white}{\star\star\star}}$ & \scriptsize (2.086)$^{\textcolor{white}{\star\star\star}}$ \\
    \multirow{2}[0]{3.5cm}{SCF income share of the top $k$\%} & 0.817$^{\textcolor{black}{\star\star\star}}$ & 0.772$^{\textcolor{black}{\star\star\star}}$ & 0.763$^{\textcolor{black}{\star\star\star}}$ & 0.789$^{\textcolor{black}{\star\star\star}}$ & 0.633$^{\textcolor{white}{\star\star\star}}$ & 0.321$^{\textcolor{white}{\star\star\star}}$ \\
          & \scriptsize (8.204)$^{\textcolor{white}{\star\star\star}}$ & \scriptsize (7.799)$^{\textcolor{white}{\star\star\star}}$ & \scriptsize (5.968)$^{\textcolor{white}{\star\star\star}}$ & \scriptsize (3.837)$^{\textcolor{white}{\star\star\star}}$ & \scriptsize (1.673)$^{\textcolor{white}{\star\star\star}}$ & \scriptsize (0.727)$^{\textcolor{white}{\star\star\star}}$ \\\midrule
    $R^2$    & 0.918 & 0.910 & 0.856 & 0.710 & 0.318 & 0.081 \\\bottomrule
    \end{tabular}}%
  \label{tab:income_estimation_results_regression_PUF_on_SCF}%
   \\\begin{tabular}{@{}c@{}} 
\multicolumn{1}{p{\columnwidth -0.65in}}{\footnotesize  This table summarizes estimation results from six unweighted linear regressions of the PUF income shares on a constant and the SCF income shares. I report estimated coefficients and $t$-statistics in parentheses. ``$^{\star\star\star}$'' denotes statistical significance at the 99 percent significance level.}
\end{tabular} 
\end{table}%

\subsection{Key findings}
\label{subsec:income_Key findings}
A natural question is which data set, the SCF survey data or the PUF sample of tax records, proves more reliable in analyzing top income inequality? My study suggests that when interested primarily in estimates of top income shares, the PUF is better than the SCF. However, when interested in top income inequality more broadly defined, the answer depends on the research question at hand.

The main advantage of using the PUF lies in higher data frequency and more precise estimates.\footnote{Moreover, as emphasized by \citet{atkinson_2011_topincomes}, tax data are available for longer time horizons and many more countries than are survey data, which makes it possible to study structural shifts in income distributions spanning several decades and to conduct cross-country  comparisons.} As discussed in detail in Section \ref{subsec:Income_analysis_Relative magnitudes of sampling error across data sets}, the PUF sampling error is five times smaller than that of the SCF. Moreover, the SCF imputation error, briefly characterized in Section \ref{subsec:income_Standard error decomposition}, introduces an additional and for the earlier years non-negligible layer of uncertainty to the SCF point estimates, whereas the PUF nonresponse error is likely to be inconsequential. Consequently, once all sources of error are accounted for, the SCF standard error is likely to be \textit{at least} five times larger than that of the PUF. Note that though this study does not estimate all components of TSE, it provides qualitative evidence suggesting that, once accounted for, measurement error in the SCF is still likely to be much more substantial than in the PUF. 

A natural question is whether more households could be surveyed for the SCF with the objective of producing more precise estimates of income concentration. Given that the cost of conducting the 2015 SCF was equal to \$18 million, increasing the SCF sample size would be an expensive undertaking, and therefore, any such decision would require a detail cost-benefit analysis, which is beyond the scope of the current paper.

Since SCF estimates are considerably less precise than those constructed using the PUF, can they still be used to draw reliable conclusions regarding top income inequality? My study suggests that while the SCF can be used to answer most questions regarding the long-term dynamics in less granular income shares, the data is not well-suited to analyze short-term horizons or year-to-year changes. 

Regarding long-term dynamics, in Section \ref{subsec:income_Long-term trend in income inequality}, I find that both the SCF and the PUF suggest a statistically significant increase in all of the six top-decile income shares under consideration. However, as we move from less to more granular income shares, the accuracy of the SCF point estimates deteriorates (see Section \ref{subsec:Income_analysis_Relative magnitudes of sampling error across data sets}).  Moreover, as discussed in further detail in Section \ref{subsec:income_Relative magnitudes of sampling error across income shares}, relative increments in CVs in the SCF are much more pronounced than those in the PUF.  Consequently, using the SCF for the estimation of more granular income shares leads to a greater loss in precision than if we were to use the PUF. Lastly, Section \ref{subsec:number_of_observations} shows that the (weighted) number of observations in the SCF above the 99.9 and 99.99 income fractiles is insufficient in both absolute and relative
terms, especially when compared to the large number of observations in the  
PUF. Therefore, even though the SCF may be useful for analyzing long-term dynamics in income shares of the top 10 or 5 percent, it should not be used in studies that focus on the top 0.1 or 0.01 percent. 

Yet \citet{saez_2016_wealthinequality}, \citet{bricker_2018_how_much}, \citet{seaz_2020_revising_after_the_revisionsists}, and others continue to rely upon the SCF estimates of the most granular income and wealth shares of the top 0.1 and the top 0.01 percent. %Writing Center 
In particular, the estimates in question are used as reference points in assessing income and wealth concentration measures obtained using alternative data sets and/or different estimation techniques. Most importantly, such comparisons are done without accounting for the SCF standard error, which affects not only the SCF point estimates, but foremost, the short- and long-term dynamics in income and wealth inequality.  A notable exception is \citet{kopczuk_2004_top_wealth_shares} who acknowledge the small number of observations in the SCF and find the survey data unreliable for the estimation of wealth shares for groups smaller than the top 0.5 percent.
%Writing Center

Regarding short-term dynamics, I find that large confidence intervals around the SCF point estimates may falsely suggest lack of a statistically significant increase in income shares from one SCF survey year to another. For instance,  whereas the SCF does not suggest a statistically significant change in top income shares before, during, and after the 2007--2009 Great Recession, the PUF clearly illustrates an initial
drop (2006--2009) followed by a statistically significant increase (2009--2012), to at or even above
the pre-recession levels. 

On the other hand, when it is necessary to control for a wide range of demographic and socio-economic characteristics or examine the composition of top earners by age, sex, or marital status, the PUF cannot be used, and instead, one must rely upon the SCF.\footnote{Even in the highly-restricted INSOLE sample data, information on taxpayers' demographics and socio-economic characteristics is limited to age and sex.} In order to arrive at credible results using the SCF, it is important to account for the survey's small number of observations, and, as such, refrain from estimating very granular income shares or analyzing small population subgroups. For instance, while I presume the SCF to accurately  estimate the difference in trends in top income
inequality between married and unmarried households, the number of observations is likely insufficient to produce credible results on married and unmarried households with and without dependents.

An ideal data set for studying top income inequality would comprise a large number of observations and a rich set of demographic and socio-economic characteristics. Could such a data set be constructed from a merge of the data from the SCF and the PUF? While the answer to this question is beyond the scope of the current paper, as shown in Section \ref{subsec:cross-correlations}, there exists an explicit statistical link between between the PUF and the SCF, which could be used in a follow-up research project to analyze how economic factors could impact PUF through its link with SCF.\footnote{Moreover, since the PUF is released with a substantial delay when compared to the SCF (the latest available data set is from 2012), it is even more critical to statistically link the PUF and the SCF. While  naive way would be through regressions, a more sophisticated approach belongs to future research.}

Specifically, combining information from the SCF and the PUF would be particularly advantageous for studying racial and ethnic income inequality. According to the New York Times, a black family with a newborn baby has a median household income of \$36,300, which compares to \$80,000 for white households.\footnote{See \href{https://www.nytimes.com/2020/06/09/your-money/race-income-equality.html}{https://www.nytimes.com/2020/06/09/your-money/race-income-equality.html}  (accessed on October 7, 2020).} Moreover, since COVID-19 is having disproportionate impact on people of color (blacks have the highest death toll per 100,000, in comparison to whites, Asians, Latinos, and indigenous Americans)\footnote{See \href{https://www.apmresearchlab.org/covid/deaths-by-race}{https://www.apmresearchlab.org/covid/deaths-by-race}  (accessed on October 7, 2020).} we can expect these longstanding racial and ethnic disparities in income to grow. So far, the Tax Policy Center and the Joint Committee of Taxation used administrative tax records in combination with the SCF survey data to examine the impact of the 2017 federal corporate tax cut (Trump's tax) on racial and ethnic income inequality.\footnote{See \href{https://www.nytimes.com/2018/10/11/business/trump-tax-cuts-white-americans.html}{https://www.nytimes.com/2018/10/11/business/trump-tax-cuts-white-americans.html} (accessed on October 7, 2020).} Given the widening gap in income between  black and white Americans, similar analyses are called for in the context of COVID-19, the 2007--2009 Great Recession, and any other major shock to the post-WWII US economy. 

% Writing Center
Consequently, this study does not portray the SCF and the PUF as supplements but rather as complementary data sources that can and should be used interchangeably in order to best answer a specific research question. However, this is often not possible due to a strictly limited access to individual-income tax returns, which is only granted to a handful of researchers. For this reason, this paper advocates for a broader access to various sources of administrative micro-level data (such as the PUF), supporting the calls by \citet{card_2010_expanding_access} and others.
% Writing Center
\subsubsection{Conclusion}
Regarding top income inequality, I find a statistically significant increase in all six top-decile income shares under consideration in the twenty-two-year period between 1991 and 2012. Specifically, the weighted least square regression analysis of the PUF estimates suggests an increase in income shares of the top 10, 5, 1, 0.5, 0.1, and 0.01 percent by 16, 21, 33, 40, 56, and 75 percent, respectively. These results are in line with those published in the related literature, supporting the claim of the long-term trend toward growing income concentration.

\section{Empirical results on wealth shares} 
\label{sec:wealth_empirical_results}
My analysis of the wealth shares within the top 10 percent is based on
four sets of estimates: one constructed using the SCF and three constructed using the PUF.  The first set of PUF estimates corresponds to wealth shares estimated under a homogeneity assumption imposed on all rates of return of the underlying capitalization model, whereas the other two sets allow for heterogeneous rates of return on taxable interest-bearing assets. In Section \ref{subsec:wealth_number_of_observations}, I examine the number of observations available for the estimation of top wealth inequality in the SCF and the PUF. In Sections \ref{subsec:wealth_poin_estiamte}--\ref{subsec:Wealth inequality and the Great Recession}, I  discuss point estimates, standard errors, and the long- and short-term dynamics in top wealth inequality. Finally, Section \ref{subsec:wealth_key_findings} summarizes my main finding and concludes.

\subsection{Number of observations}
\label{subsec:wealth_number_of_observations}
I start my analysis with a brief description of the total and weighted number of observations available for the estimation of the six top-decile wealth shares under consideration in the SCF and the PUF between 1991 and 2012. Since neither of the two data sets contain any missing values, the number of observations available to study wealth is equal to the total number of observations in each of the two data sets.
\begin{table}[htbp]
  \centering
  \caption{Ratio of the SCF number of observations to the PUF number of observations used in the estimation of the wealth shares within the top 10 percent}
    \resizebox{.8\textwidth}{!}{\begin{tabular}{C{1.5cm}C{3.0cm}C{1.5cm}C{1.5cm}C{1.5cm}C{1.5cm}C{1.5cm}C{1.5cm}} \toprule
    \multicolumn{1}{c}{\multirow{2}[0]{*}{\textbf{Year}}} & \multicolumn{1}{c}{\multirow{2}[0]{3.0cm}{\textbf{Unweighted: total sample}}} & \multicolumn{6}{c}{\textbf{Weighted: above the \boldmath$k$ income fractile}} \\
    \cline{3-8}
          &       & \textbf{90} & \textbf{95} & \textbf{99} & \textbf{99.5} & \textbf{99.9} & \textbf{99.99} \\ \midrule
    1991  & 3.4   &             2.2  &            2.1  &            1.8  &            1.8  &            1.9  &            2.4  \\
    1994  & 4.5   &             3.0  &            2.8  &            2.2  &            2.0  &            1.8  &            2.8  \\
    1997  & 3.9   &             2.3  &            2.2  &            1.6  &            1.5  &            1.5  &            2.1  \\
    2000  & 3.0     &             1.6  &            1.5  &            1.1  &            1.1  &            1.3  &            2.2  \\
    2003  & 3.4   &             1.9  &            1.8  &            1.5  &            1.5  &            1.7  &            3.1  \\
    2006  & 3.0     &             2.0  &            1.9  &            1.7  &            1.8  &            3.1  &            7.0  \\
    2009  & 4.2   &             2.6  &            2.3  &            1.9  &            2.0  &            2.9  &            7.7  \\
    2012  & 3.5   &             2.0  &            1.8  &            1.5  &            1.7  &            2.7  &            7.1  \\\bottomrule
    \end{tabular}}%
  \label{tab:wealth_weighted_number_of_obsrevations}%
  \\\begin{tabular}{@{}c@{}} 
\multicolumn{1}{p{\columnwidth -1.4in}}{\footnotesize  Ratios are expressed in percent}
\end{tabular} 
\end{table}%

 Consequently, as indicated in Table \ref{tab:wealth_weighted_number_of_obsrevations}, there are, on average, 30 times more observations available in the PUF than there are in the SCF. 
 
Large discrepancies also exist between the SCF and the PUF with respect to the weighted number of observations above high-order fractiles. In particular, even though the SCF oversamples wealthy households, the weighted number of observations in the far right tail of wealth distribution is much smaller in the SCF than in the PUF. Consequently, as discussed in more detail in Section \ref{subsec:wealth_relative_magnitudes_of_sampling_error_across_data_sets}, the SCF confidence intervals are substantially wider than those constructed using the PUF.

\subsection{Point estimates}
\label{subsec:wealth_poin_estiamte}
\begin{figure}[hbt] 
\centering
\includegraphics[scale=0.42]{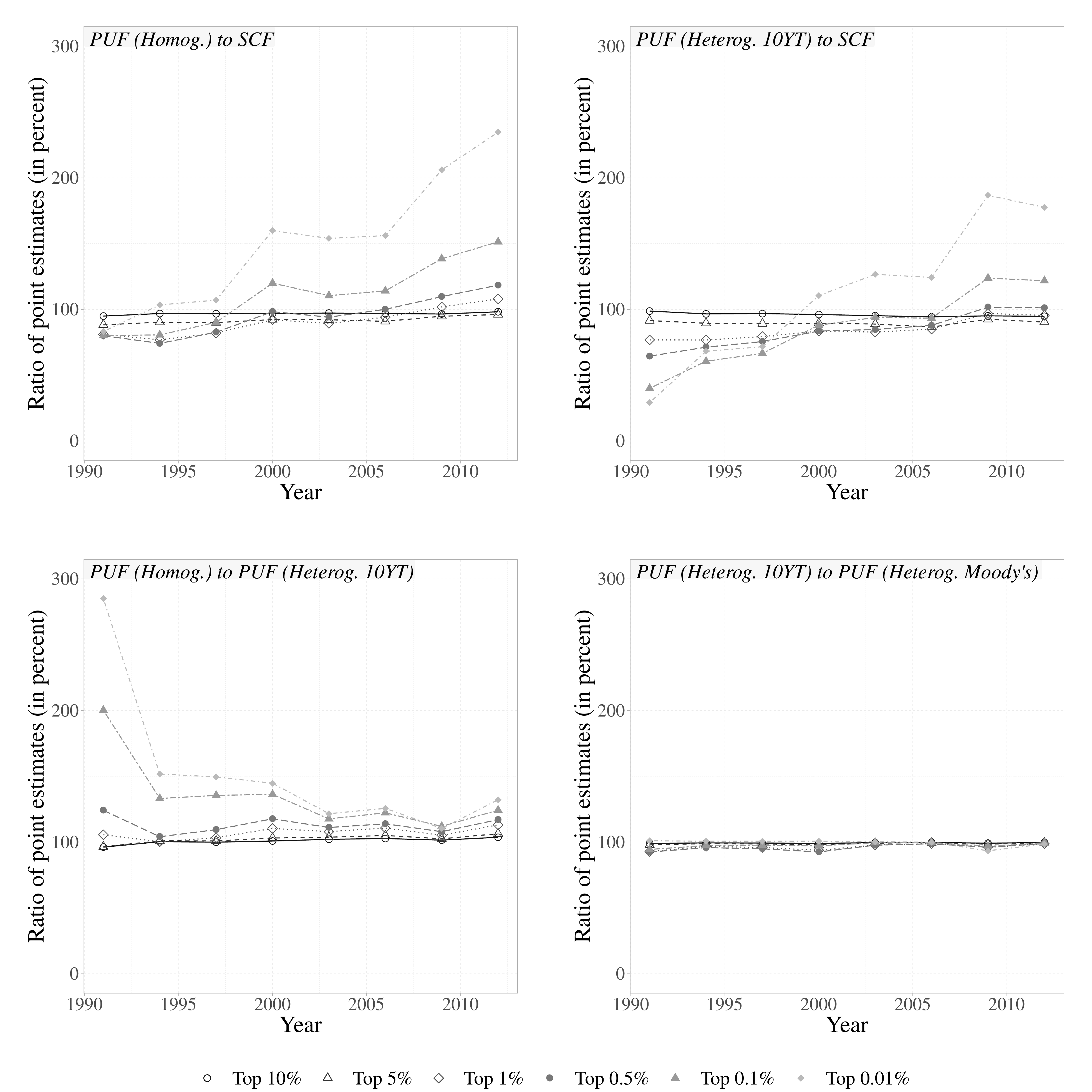}
\caption{Ratio of point estimates for the wealth shares within the top 10 percent}
\label{fig:wealth_point_estimates}
\end{figure}

My second set of results pertains to point estimates of the six wealth shares of the top 10 to the top 0.01 percent between 1991 and 2012. As indicated in Figure \ref{fig:wealth_point_estimates}, I find that the SCF and the PUF often disagree with respect to levels of the estimated wealth shares, with the SCF implying larger shares of the top 10, 5, 1, and 0.5 percent, and smaller shares of the top 0.1 and 0.01 percent. 
Since
at the moment, I do not estimate direct benefit pensions using the SCF, it is beyond the scope of the present paper to determine whether some of the observed discrepancies in point estimates could be explained by differences in the operational definitions of wealth between the SCF
and the PUF. 
\begin{figure}[hbt]
\centering
\includegraphics[scale=0.42]{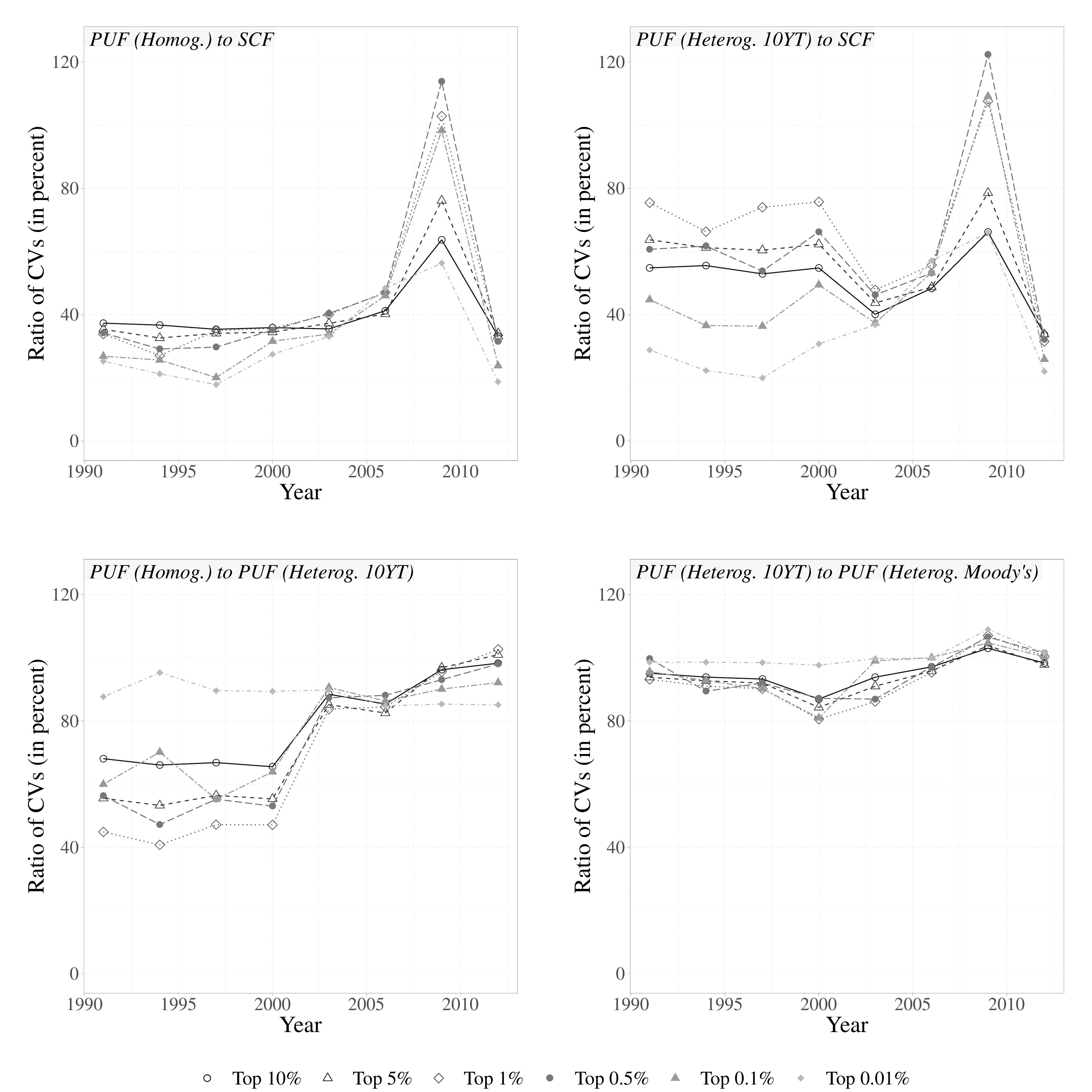}
\caption{Ratio of CVs for the wealth shares within the top 10 percent}
\label{fig:wealth_ratios_of_CV}
\end{figure}

In addition to comparing the SCF and the PUF, I analyze ratios between the three different sets of PUF estimates. Whereas I do not find significant differences in the estimated wealth shares between the two heterogeneous sets of PUF estimates, I observe non-trivial discrepancies between those estimated using homogeneous and heterogeneous models. For less granular income shares, the observed discrepancies are small to moderate. However, for more granular wealth shares of the top 0.1 and 0.01 percent, the differences become large, especially over a ten-year period between the early 1990s and the early 2000s. For example, whereas the homogeneous model suggests an increase in wealth shares of the top 0.01 percent from 3.2 to 10.2 percent between 1991 and 2012, the heterogeneous model indicates an increase from 1.25 to 7.7. Overall, my analysis suggests that the differences between the homogeneous and heterogeneous PUF estimates are similar in magnitude to those observed between the SCF and the PUF. Consequently, except for wealth shares of the top 10 percent, there is little agreement between the different set of estimates, which makes the analysis of wealth inequality considerably more challenging than that of income.

\subsection{Relative magnitudes of sampling error across data sets}
\label{subsec:wealth_relative_magnitudes_of_sampling_error_across_data_sets}
In the present section, I focus on sampling errors. In particular, I analyze ratios of CVs computed for the six wealth shares of the top 10 to the top 0.01 percent between 1991 and 2012.  Like for top income inequality, I find sampling error in the SCF to be considerably bigger than that in the PUF---a result of the great discrepancy in the number of observations between the two data sets.  As indicated in Figure \ref{fig:wealth_ratios_of_CV},  CVs for the SCF are 20--80 percent larger than those for the PUF. Therefore, if we were only concerned with the precision of the estimates, the PUF estimates would be preferred. However, in order to determine the most suitable set of estimates it is necessary to consider a wider range of factors, including estimates' accuracy and their possible dependence on wrong modeling assumptions.

\subsection{SCF standard error decomposition}
\label{subsec:wealth_scf_standard_error_decomposition}
In Table \ref{tab:wealth_error_decomposition}, I present ratios of sampling error to  total standard error for the six wealth shares of the top 10 to the top 0.01 percent between 1988 and 2018. My analysis suggests that, unlike for income, sampling error in the SCF estimates of top wealth inequality does not constitute the only considerable source of variation. 

\begin{table}[hbt!]
  \centering
  \caption{Ratio of sampling error to total standard error in the SCF estimates of the wealth shares within the top 10 percent}
    \resizebox{.7\textwidth}{!}{\begin{tabular}{C{1.7cm}C{1.7cm}C{1.7cm}C{1.7cm}C{1.7cm}C{1.7cm}C{1.7cm}} \toprule
    \multicolumn{1}{c}{\multirow{2}[0]{*}{\textbf{Year}}} & \multicolumn{6}{c}{\textbf{Wealth share of the top \boldmath$k$\%}} \\
        \arrayrulecolor{cadetgrey}\cline{2-7}
          & \textbf{10\%} & \textbf{5\%} & \textbf{1\%} & \textbf{0.5\%} & \textbf{0.1\%} & \textbf{0.01\%} \\\midrule
    1988  & 68.58 & 78.39 & 56.34 & 58.96 & 54.88 & 73.84 \\
    1991  & 40.34 & 42.81 & 58.10 & 39.70 & 56.92 & 92.79 \\
    1994  & 88.12 & 88.52 & 85.23 & 70.45 & 72.11 & 90.13 \\
    1997  & 42.01 & 52.23 & 36.29 & 36.44 & 56.31 & 82.07 \\
    2000  & 55.68 & 66.88 & 71.76 & 73.73 & 77.30 & 89.19 \\
    2003  & 36.97 & 46.38 & 81.59 & 86.61 & 85.27 & 93.68 \\
    2006  & 67.56 & 70.79 & 66.08 & 72.78 & 96.16 & 96.21 \\
    2009  & 55.23 & 78.11 & 61.52 & 47.01 & 70.72 & 96.76 \\
    2012  & 91.24 & 87.25 & 95.82 & 97.29 & 94.45 & 95.14 \\
    2015  & 68.68 & 79.05 & 91.59 & 82.17 & 91.86 & 84.98 \\
    2018  & 85.3  & 65.3  & 76.5  & 83.1  & 93.0  & 95.7 \\\bottomrule
    \end{tabular}}%
  \label{tab:wealth_error_decomposition}%
   \\\begin{tabular}{@{}c@{}} 
\multicolumn{1}{p{\columnwidth -2in}}{\footnotesize  Ratios are expressed in percent}
\end{tabular} 
\end{table}%

In particular, I find that between 1988 and 2018, imputation error often accounted for between 30 and 50 percent of total standard error in the estimated wealth shares. From the perspective of policy-makers, this result suggests that revising the SCF imputation procedure and/or introducing new interview techniques aimed at mitigating item nonresponse could prove effective in reducing the uncertainty in the SCF estimates of top-wealth inequality. This is an important finding, since reducing the SCF sampling error by significantly increasing the number of households selected for the survey is nearly impossible due to the large costs and organizational complexity associated with conducting the survey.
\begin{figure}[H]
\centering
\includegraphics[scale=0.85]{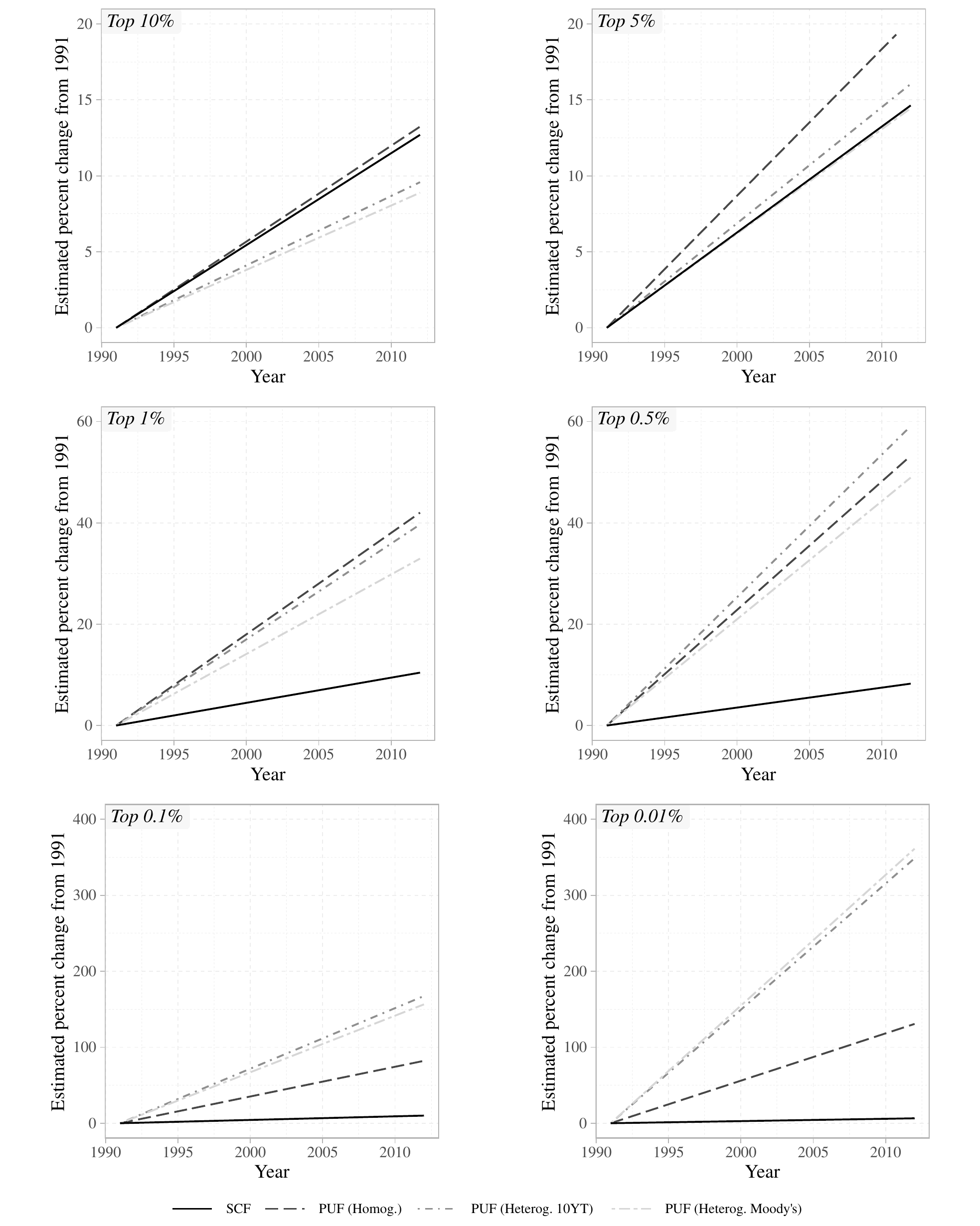}
\caption{Estimated increase in the wealth shares within the top 10 percent in the SCF and the PUF}
\label{fig:wealth_compact_trends_long}
\end{figure}

\subsection{Long-term dynamics}
\label{subsec:wealth_long-term_dynamics}
In order to analyze long-term dynamics in wealth inequality between 1991 and 2012, I estimate weighted linear regressions of top-decile wealth shares on a constant and linear time trend. As illustrated in Figure  \ref{fig:wealth_compact_trends_long}, the SCF regression results support the claim of  a statistically significant increase in three out of the six wealth shares under consideration (see Table C.6 in Appendix C of the Online Supplementary Material for the estimation details). I find a statistically significant increase in the wealth shares of the top 10 and the top 5 percent at the 1 percent significance level, and in the wealth shares of the top 1 percent at the 10 percent significance level. The PUF estimates, on the other hand, suggest a statistically significant increase in all of the six top-decile wealth shares under consideration, irrespective of the assumption imposed on the rate of return of taxable interest-bearing assets. However, there exist considerable differences in the estimated trend lines between the homogeneous and heterogeneous sets of estimates. For the wealth shares of the top 10 and 5 percent, the homogeneous model suggests a greater increase in inequality between 1991 and 2012 than the heterogeneous model. Yet  for the wealth shares of the top 0.1 and 0.01 percent the opposite is true: the heterogeneous model indicates a much more pronounced increase than the homogeneous model. In particular, the estimated percent change in the wealth shares of the top 0.01 percent from 1991 to 2012 exceeds 300 percent for the heterogeneous model while being equal to over 100 percent for the homogeneous  model. Therefore, without taking a strong stance on what model better  describes the state of the economy, the PUF estimates remain inconclusive with respect to the magnitude by which the top wealth inequality has increased.

\subsection{Wealth inequality and the Great Recession}
\label{subsec:Wealth inequality and the Great Recession}
In the present section, I discuss the impact of the 2007--2009 Great Recession on wealth inequality, with the main focus on wealth shares of the top 10 percent between 2006 and 2012. 
\begin{figure}[htb!]
\centering
\includegraphics[scale=0.55]{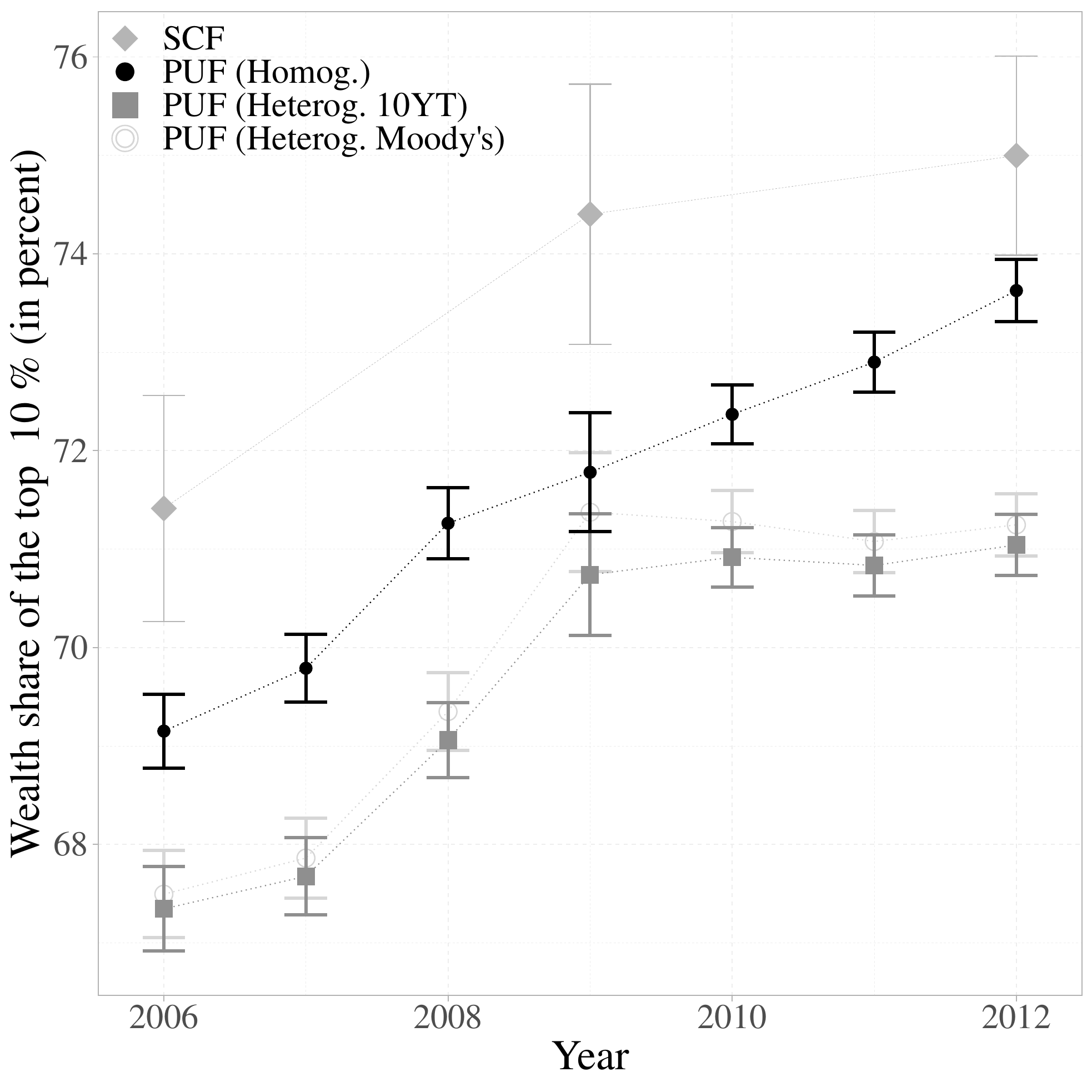}
\caption{Wealth shares of the top 10 percent before and after the 2007--2009 Great Recession measured in the SCF and the PUF. Error bars indicate 95 percent confidence intervals around point estimates}
\label{fig:wealth_great_recession}
\end{figure}
As illustrated in Figure \ref{fig:wealth_great_recession}, I find considerable differences between the homogeneous and heterogeneous PUF estimates. Specifically, whereas homogeneous estimates had been steadily increasing in the aftermath of the Great Recession, heterogeneous estimates came to near standstill once the recession ended. In other words, the two sets of estimates lead to drastically different conclusions regarding the short-term dynamics in top wealth inequality. One set of estimates suggests a statistically significant increase of 2 percentage points between 2009 and 2012, whereas the other set implies no statistically significant change since the recession came to an end in 2009. Therefore, with reference to Section \ref{subsec:wealth_long-term_dynamics}, I find the PUF estimates inconclusive not only  in relation to the long-term dynamics in top wealth inequality, but also when examining shorter-time horizons and times of rapid economic changes, such as the 2007--2009 Great Recession.

\subsection{Key findings}
\label{subsec:wealth_key_findings}
My analysis suggests that the PUF estimates of top wealth inequality are largely inconclusive---a result of substantial differences in the estimates resulting from the set of assumptions imposed on the underlying capitalization model. 

As indicated in Section \ref{subsec:wealth_poin_estiamte}, I find non-trivial differences in the PUF point estimates obtained under the homogeneity assumption and those computed using heterogeneous rates of return on taxable interest-bearing assets. In addition to disparities in point estimates, the two sets of estimates  differ with respect to short- and long-run trends in top wealth inequality. First, regarding the long-term dynamics,  the estimated trend lines lead to different conclusions regarding the severity of the ongoing crisis linked to rising inequality. Second, regarding the short-term dynamics, I find that whereas the homogeneous  estimates support the claim of an increase in wealth inequality following the 2007--2009 Great Recession, the heterogeneous estimates do not indicate a statistically significant change.

Assuming different rates of return leads to considerably different dynamics in top wealth inequality. Then, since the assumptions matter, a natural question is what rates of return should we be considering for each of the asset classes under consideration?  Should they be mostly homo- or heterogeneous? Should we allow them to vary by income or wealth percentile?  Should they depend on portfolio composition or regional macroeconomic conditions?  In this paper,  following \citet{bricker_2018_how_much}, I consider a heterogeneous rate of return on only one class of assets, where I impose a hard, and rather unrealistic, cut-off between high and low rates of return. Therefore, another important question is whether a different and more realistic set of assumptions would result in distinct estimates of top wealth inequality. Based on the extent to which the PUF estimates obtained under the homogeneity assumption and those computed using heterogeneous rates of return on taxable interest-bearing assets differ, I presume the resulting estimates would be considerably different from those currently obtained. This presumption is line with \citet{kopczuk_2004_top_wealth_shares}, who, among many others, expressed concerns about the estimation of wealth using tax-based income data, resulting from ``substantial and unobservable heterogeneity in the returns of many assets, especially corporate stock.''

Since the PUF estimates of top wealth inequality are functions of numerous assumptions imposed on the underlying capitalization model, is the SCF a better alternative? My analysis suggests that for the less granular income shares of the top 10 to the top 0.5 percent the SCF appears more reliable. This results mainly from the fact that the SCF \textit{measures} wealth, whereas the PUF \textit{infers} wealth from data on income. However, using the SCF for the estimation of top wealth shares has its own limitations resulting from a small number of observations and large confidence intervals. Therefore, whereas the SCF can be used effectively to analyze the less granular wealth shares of the top 10 to the top 0.5 percent, the data remain inadequate to get a realistic picture of the wealth shares of the top 0.1 and 0.01 percent.\footnote{The wealth shares of the top 10 to the top 0.5 percent constructed using the SCF can be further refined by adding the wealth of the Forbes 400 wealthiest Americans and the value of direct benefit pensions as in \citet{bricker_2016_measuring}.}  

It is important to note that while resolving the problem of the large standard errors in the SCF would require making changes to the survey design and/or the SCF imputation procedure, future research could focus on refining the \citet{saez_2016_wealthinequality}'s capitalization model, with the objective of producing more reliable estimates of wealth inequality using the PUF. 
A significant research effort has been already undertaken by  \citet{smith_2019_top_wealth_in_america} and \citet{seaz_2020_comments_on_smith_zidar_zwick, seaz_2020_revising_after_the_revisionsists}, adding to our understanding of the numerous advantages and disadvantages embedded in studying wealth inequality using capitalization methods.

Since both the SCF survey data and the PUF individual-income tax returns pose major challenges for the estimation of top wealth inequality in the US, this paper supports \citet{seaz_2020_revising_after_the_revisionsists} in calling for ``more and improved statistics on inequality.'' The authors further emphasize that ``we could and should do better to measure US wealth inequality than rely on a triennial survey of 6,200 families (the Survey of Consumer Finances) or indirectly infer asset ownership based on income flows (the capitalization method).''

\subsubsection{Conclusions}
Regarding the twenty-two year period between 1991 and 2012, my analysis suggests a statistically significant increase in three out of the six wealth shares under consideration. Specifically, the weighted least square regression analysis of the SCF estimates suggests a statistically significant increase in wealth shares of the top 10, 5 and 1 percent by 8.4, 8.0, and 3.3 percentage points, respectively, and an insignificant increase in wealth shares of the top 0.5 percent. Since, as indicated above, neither the SCF nor the PUF proves credible to analyze more granular wealth shares of the top 0.1 and 0.01 percent,  this study does not draw any conclusions related to top wealth inequality above the 99.5 wealth fractile. Lastly, my study does not support \citet{saez_2016_wealthinequality}'s conclusion 
related to a decline in the wealth shares of the top 10 less the top 1 percent (i.e., those with wealth above the 90 wealth percentile but below the 99 wealth percentile). Instead, I find that the estimated change in the wealth shares of the top 10 percent between 1991 and 2012 exceeds the estimated change in the wealth shares of the top 1 percent by 2.5 percentage points, a finding that contradicts the claim of rising wealth inequality among the richest.
%%%%%%%%%%%%%%%%%%%%%%%%%%%%%%%%%%%%%%%%%%%%%%%%%%%%%%%%%%%%%%%%%%%%%%%%%%%
\section{Structural exercise}
\label{sec:structural_exercise}
In the following, I investigate how the SCF and the PUF data-driven errors in estimates of top income inequality affect outcomes of structural macroeconomic models. In Section \ref{subsec:structural_model}, I introduce the theoretical foundation of my study, which is the augmented random growth model with type-dependence proposed by \citet{gabaix_2016_dynamics}. Next, in Section \ref{subsec:structural_calibration}, I discuss details of my calibration strategy that, in contrast to the default approach, involves multiple model calibrations, each time to a different value of the calibration target randomly drawn from the estimated 95 percent confidence interval. Finally, in Section \ref{subsec:results}, I present the results of my structural exercise,  and  conclude on how data-driven errors in calibration targets may affect outcomes of macroeconomic and policy-oriented studies more generally.

\subsection{Model}
\label{subsec:structural_model}
Consider a continuum of workers, where each worker $i$ is either high- or low-type, with high-type workers having higher mean growth rate of income than low-type workers. Workers enter the labor market as high-types with probability $\theta$ and as low-types with probability $1-\theta$. Whereas no worker of low-type can ever become a high-type, high-type workers do switch to a low-type  with probability $\alpha$. Since a low-type is an absorbing state, high-type workers can switch to a low-type at most once in their life-time. Moreover, workers retire at rate $\delta$ and are replaced by new labor entrants with wages drawn from a known distribution $\psi$.

Next, let $x_{it}$ denote a natural logarithm of income of worker $i$ of type $j$ at time $t$, and let the dynamics of $x_{it}$  be given by a type-dependent random growth model as in \citet{gabaix_2016_dynamics}:
\begin{align}
dx_{it} = \mu_jdt + \sigma_jdZ_{it} + \text{Injection} - \text{Death},\quad j=\left\lbrace H,L\right\rbrace,
\end{align}
where $Z_{it}$ is a standard Brownian motion, $\mu_j$ and $\sigma_j^2$ are type-dependent mean and variance of growth rate of log income, and where $H$ and $L$ are shorthand notations for high- and low-types, respectively.

Moreover, assume that the stationary distribution of log income has a Pareto tail,
\begin{align}
\mathbb{P}\left(x_{it}>x\right) \sim C e^{-\xi x},
\end{align}
where $C$ is a constant and $\xi>0$ is a power law exponent given by 
\begin{align}
\xi = \frac{-\mu_H+\sqrt{\mu_H^2 + 2\sigma_H^2\left(\delta-\alpha\right)}}{\sigma_H^2}.
\end{align}

Finally, impose that the economy is in a Pareto steady state with $\sigma_H=0.15$, $\alpha=1/6$, and $\delta=1/30$, and assume that $\mu_H$ is calibrated from a one-to-one-mapping between the inverse of the power law exponent, say $\eta$, and the empirical ratio of two, top-decile  income shares:
\begin{align}
\eta=\frac{1}{\xi}=1+\log_{10}\left(\frac{p_{(k/10)}}{p_k}\right).
\end{align}
%%%%%%%%%%%%%%%%%%%%%%%%%%%%%%%%%%%%%%%%%%%%%%%%%%%%%%%%%%%%%%%%%%%%%%%%%%%%
\subsection{Calibration}
\label{subsec:structural_calibration}
Unlike \citet{gabaix_2016_dynamics}, where the authors consider a single value of the inverse of the power law exponent, $\eta$, I calibrate the model to multiple values of $\eta$ drawn from a 95 percent confidence interval around the $\eta$ point estimate. In particular, I conduct $B=100$ random draws, where for each drawn value of $\eta$, I solve the model for the transition dynamics to a new steady state.

\citet{gabaix_2016_dynamics} compute a point estimate of $\eta$ using data on top income shares from the World Income Database (WID), whereas this paper's focus is on the SCF and the PUF. Moreover, the authors calibrate the model to the 1973 WID, whereas the SCF and the PUF data considered in this analysis start in 1988 and 1991, respectively.  An obvious solution to this problem would be to estimate $\eta$ using the 1988 SCF and/or the 1991 PUF. However, this strategy would not allow me to directly compare my results to those in  \citet{gabaix_2016_dynamics}, since I would be effectively investigating transition dynamics over a different time horizon: 1988--2065 (or 1991--2068) versus 1973--2050. Moreover, it would likely require me to consider a different magnitude of the shock and result in changes to the model parameters describing the initial Pareto steady state. Instead, I directly build upon the exercise in \citet{gabaix_2016_dynamics}  at the cost of making two assumptions regarding hypothetical values of the SCF and the PUF estimates of $\eta$ in 1973. 

\textit{Assumption 1}: Let $\hat{\eta}_{\text{SCF}, 1973}$ and $\hat{\eta}_{\text{PUF}, 1973}$ denote the SCF and the PUF point estimates of $\eta$ in 1973, and assume that  $\hat{\eta}_{\text{SCF}, 1973}$ and $\hat{\eta}_{\text{PUF}, 1973}$ are equal to the \citet{gabaix_2016_dynamics}'s point estimate of $\eta$ obtained using the WID, say $\hat{\eta}_{\text{WID}, 1973}$:
\begin{align}
\hat{\eta}_{\text{SCF}, 1973} = \hat{\eta}_{\text{PUF}, 1973} = \hat{\eta}_{\text{WID}, 1973}=0.39.
\end{align}

\textit{Assumption 2}: Let $\text{CV}_{\text{SCF}, 1973}$ and $\text{CV}_{\text{SCF}, 1988}$ denote CVs for the SCF estimates of $\eta$ in 1973 and 1988, and let $\text{CV}_{\text{PUF}, 1973}$ and $\text{CV}_{\text{PUF}, 1991}$ denote CVs for the PUF estimates of $\eta$ in 1973 and 1991. In the following, I assume the same CVs for the SCF and the PUF point estimates of $\eta$ between the early 1970s and the early 1990s,
\begin{align}
\text{CV}_{\text{SCF}, 1973} = \text{CV}_{\text{SCF}, 1988} \quad \text{and} \quad 
\text{CV}_{\text{PUF}, 1973} = \text{CV}_{\text{PUF}, 1991}.
\end{align}

Whereas the first assumption guarantees a direct comparison with \citet{gabaix_2016_dynamics}, the second conjecture  provides me with conservative estimates of the SCF and the PUF standard errors in $\hat{\eta}_{1973}$.\footnote{The estimates are conservative since it is likely that the standard errors in the SCF and the PUF estimates of $\eta$ would have substantially decreased over the twenty-year period between the early 1970s and the early 1990s.} 

In the first step, I compute the point estimates and standard errors of $\eta$ using the 1988 SCF and the 1991 PUF. For the SCF, I find $\hat{\eta}_{\text{SCF},1988}=0.56$ and $\text{SE}\left(\hat{\eta}_{\text{SCF},1988}\right)=0.03$, which results in $\text{CV}_{\text{SCF},1988}=0.054$; for the PUF, I find $\hat{\eta}_{\text{PUF},1991}=0.51$ and $\text{SE}\left(\hat{\eta}_{\text{PUF},1991}\right)=0.001$, which results in $\text{CV}_{\text{PUF},1991}=0.003$. 

In the second step, I use Assumptions 1 and 2 to compute the standard errors in the SCF and the PUF estimates of $\eta$ in 1973: 
\begin{align}
\text{SE}\left(\hat{\eta}_{\text{SCF},1973}\right) = \text{CV}_{\text{SCF},1973} \times \hat{\eta}_{\text{SCF}, 1973} = 0.023
\end{align}
\begin{align}
\text{SE}\left(\hat{\eta}_{\text{PUF},1973}\right) = \text{CV}_{\text{PUF},1973} \times \hat{\eta}_{\text{PUF}, 1973}=0.001.
\end{align}

Having estimated $\text{SE}\left(\hat{\eta}_{\text{SCF},1973}\right)$ and $\text{SE}\left(\hat{\eta}_{\text{PUF},1973}\right)$, I construct 95 percent confidence intervals around $\hat{\eta}_{\text{SCF},1973}$ and $\hat{\eta}_{\text{PUF},1973}$. For the SCF, the 95 percent confidence interval is given by  $\left[0.36, 0.45\right]$; for the PUF, the 95 percent confidence interval is given by $\left[0.40, 0.41\right]$.

In the third and final step of my analysis, I conduct $B=100$ random draws from $\left[0.36, 0.45\right]$ and $\left[0.40, 0.41\right]$, where for each drawn value of $\eta$, I solve for the model's transition dynamics to a new steady state. For each of the two data sets, this exercise results in $B=100$ different transition dynamics to a new steady state, where the observed variability in the model's outcomes is driven solely by the underlying uncertainty in the estimate of $\eta$. 

In addition to the uncertainty in the estimate of $\eta$, the model is subject to errors arising from inaccurately assumed values of the remaining model parameters. In the following, I focus on  $\sigma_H$, which \citet{gabaix_2016_dynamics} set equal to 0.15, while pointing out that  $\sigma_H=0.15$ is a conservative estimate since ``the growth rates of parts of the population may be much more volatile (think of startups).'' While an ideal way to account for this additional layer of uncertainty would be to estimate $\sigma_H$ from the data and, then calibrate the model to a 95 percent confidence interval around the point estimate of $\sigma_H$, I leave this approach for future research. In this paper, I consider a simpler exercise, in which I assume a set of possible values of $\sigma_H$ given by $\left\lbrace 0.15, 0.175, 0.2\right\rbrace$. Note that this set is constructed in accordance with the \citet{gabaix_2016_dynamics}'s presumption that the growth rate of log income among high types is likely to exceed 0.15. Then, for each value of $\sigma_H$ and for each of the two data sets under consideration, I calibrate the model $B=100$ times, each time using a different value of $\eta$ randomly drawn from a 95 percent confidence interval  around the $\eta$'s point estimate. This approach allows me to construct a ``feasible region'' of the model's transition dynamics to a new steady state while incorporating uncertainties arising from data-driven errors in two out of the six model's parameters.

\subsection{Results}
\label{subsec:results}
In the following, I discuss two sets of results. First, I analyze the transition dynamics accounting only for the variation in the calibration target $\eta$. Then, I introduce an additional layer of uncertainty and repeat my analysis for different values of $\sigma_H$, a parameter that governs the volatility of the growth rate of log income among high types.

\subsubsection{Varying $\eta$}
My first set of results shows that data-driven errors in estimates of key macroeconomic aggregates impact outcomes of not only empirical but also structural analysis. Errors in calibration targets are carried over through the model and come to affect all outcomes of interest, including transition dynamics and the speed of convergence to a new steady state. Therefore, I find that having precise estimates of calibration targets critical for producing precise outcomes of structural analysis. 
\begin{figure}[htb]
\centering
\includegraphics[scale=0.85]{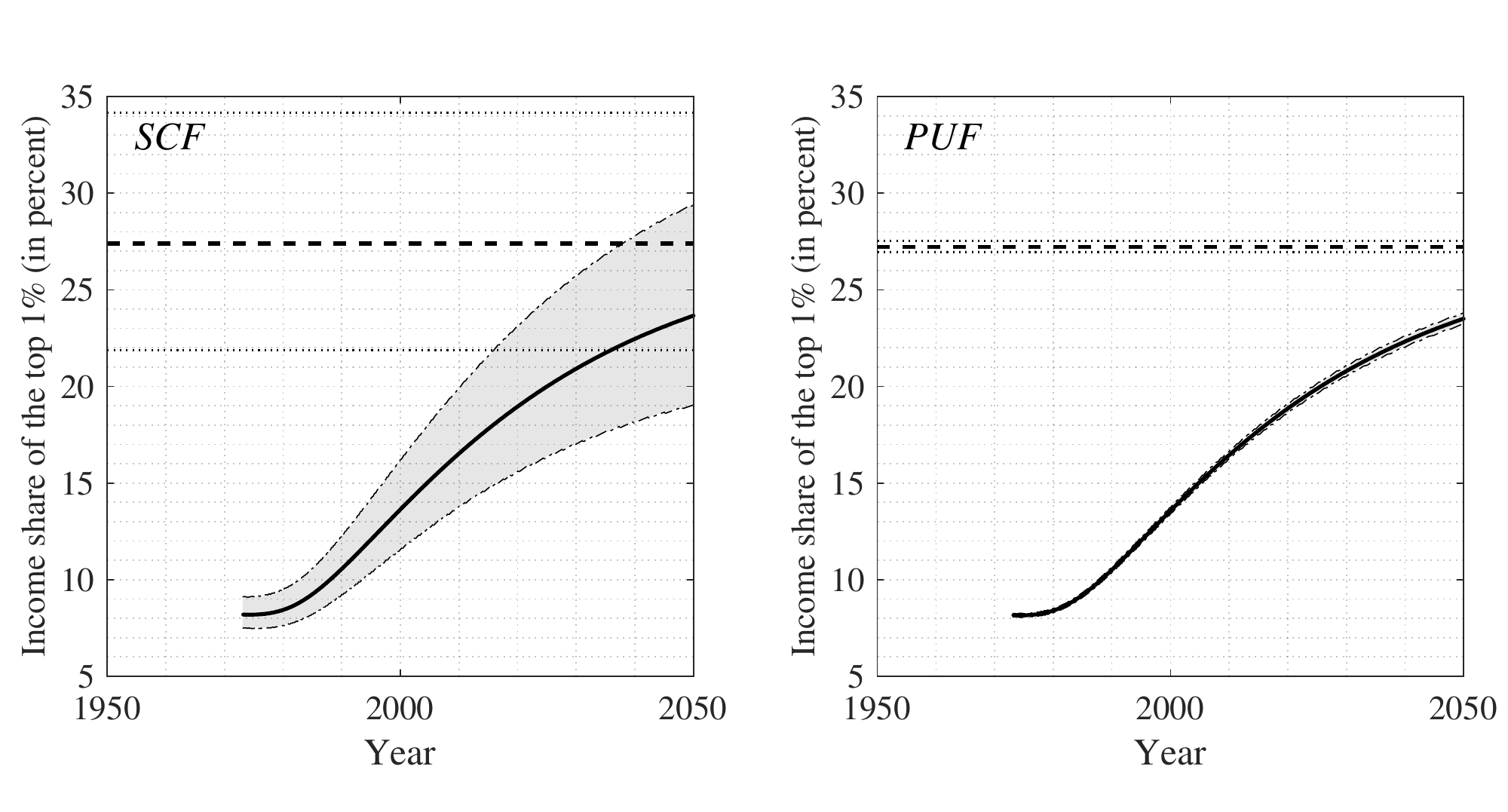}
\vspace{-1cm}
\caption{Transition dynamics, varying $\eta$}
\label{fig:structural_exercise_result}
\end{figure}
This becomes evident when comparing transition dynamics to a new steady state produced by the exact same model differing, only with respect to the level of uncertainty surrounding a single calibration target. As indicated in Figure \ref{fig:structural_exercise_result}, the PUF estimate of $\eta$ has a negligible standard error, which results in a precise estimation of the model's dynamics to a new steady state. The SCF estimate of $\eta$, on the other hand,   has a sizable standard error, which results in a large 95 percent confidence envelop around the estimated transition dynamics and in a highly imprecise estimate of a new steady state. 

\subsubsection{Varying $\sigma_H$}
\begin{figure}[htb]
\centering
\includegraphics[scale=0.81]{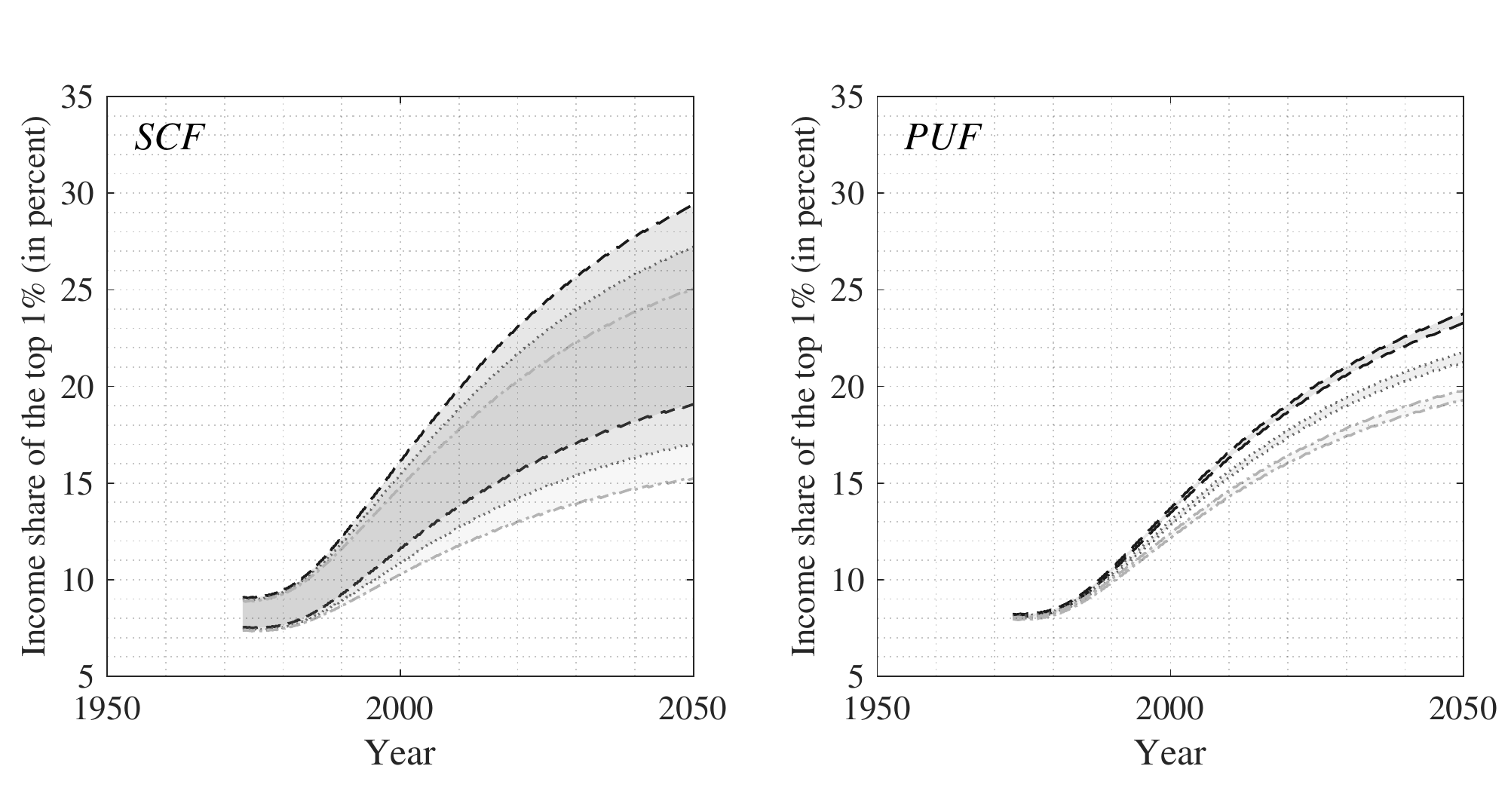}
\vspace{-1cm}
\caption{Transition dynamics, varying $\sigma_H$}
\label{fig:structural_exercise_result_sigmaH}
\end{figure}

Accounting for an additional layer of uncertainty creates an even clearer picture of the importance of conducting structural analysis using precisely estimated calibration targets. In Figure \ref{fig:structural_exercise_result_sigmaH},  for each of the three assumed values of $\sigma_H$, I present a 95 percent confidence envelope constructed  around the estimated transition dynamics to a new steady state. By construction, varying the volatility of the growth rate of log income among high types has the same effect on the model's outputs, regardless of the data set under consideration. However, when combined with the data-set-specific uncertainty in the point estimate of $\eta$, the advantage of relying upon the PUF for the model's calibration becomes evident. Using the SCF, I arrive at projections of the income shares of the top 1 percent, ranging from 15 to 30 percent as of 2050. Since such a wide range of possible values is largely uninformative, it presents no real value to policy makers evaluating proposals targeted at combating rising inequality. On the other hand, the model's projections obtained using the PUF are much more precise. The 95 percent confidence envelope around the projected income shares varies from 20 to less than 25 percent, providing policy makers with an informative rage of values, while accounting for uncertainty in two out of the six model parameters.

The above analysis is conducted using data that precede COVID-19, and does not account for the impact of the ongoing pandemic on income inequality. Therefore, Figures \ref{fig:structural_exercise_result} and \ref{fig:structural_exercise_result_sigmaH} demonstrate largely outdated projections.  This is the case since COVID-19 is likely to have far more long-lasting impacts on inequality than any other post-WWII recession, including the 2007--2009 Great Recession. Particularly, the forced shutdown of large parts of the US economy caused a dramatic spike in unemployment rates, especially among minorities and low-educated service workers in high-interaction jobs at restaurants, pubs, hotels, and entertainment venues. Moreover, unlike other parts of the world, where governments covered employees' wages for the duration of the crises, US employees were laid off without a guarantee of being re-hired. Therefore, despite the fact that measuring impact of COVID-19 on inequality would require data that are not yet available, I find it an important topic for future research, which I briefly discuss in Section \ref{sec:conclusions}.\footnote{Furthermore, the current projections do not account for recent changes to the US  tax policy, including the Trump's 2017 tax cut that lowered the corporate tax rate from 35 to 21 percent. Since such reductions in taxes are likely to widen the gap between rich and poor, extending the current analysis by accounting for not only the impact of COVID-19 but also the 2017  Trump's tax cut would deliver substantially more realistic projections that those currently obtained.}

\section{Conclusions}
\label{sec:conclusions}
This paper discusses various sources of uncertainty in studying economic inequality within and across data sets. With the focus on the six income and wealth shares of the top 10 to the top 0.01 percent, I investigate how  sampling, nonsampling, and modeling errors affect outcomes of empirical analysis conducted using the SCF and the PUF. Regarding income inequality, I find that the PUF estimates are substantially better than the SCF estimates, and consequently, whenever possible, should be relied upon for both empirical and structural analysis. On the other hand, for top wealth inequality, neither of the two data sets can be used without caution. Regarding wealth shares of the top 10 to the top 0.5 percent, I find the SCF estimates more reliable---a result of unexpected and yet-to-be-accounted-for differences among the PUF estimates that arise  from varying assumptions imposed on the underlying capitalization model. However, for the more granular wealth shares of the top 0.1 and 0.01 percent, neither of the two data sets proves credible. The PUF estimates lead to different conclusions depending on which capitalization model one applies, while the SCF estimates are unreliable as a result of the very sparse number of observations in the far right tail of wealth distribution.

In addition, using the random growth model of income from \citet{gabaix_2016_dynamics}, I illustrate how data-driven errors in calibration targets affect the outcomes of structural macroeconomic models. All in all, I find that in order for a structural analysis to be both conclusive and informative, it is necessary to rely upon precisely estimated calibration targets. As shown in Section \ref{sec:structural_exercise}, large uncertainties in the SCF estimates result in wide confidence intervals around the transition dynamics of the income shares of the top 1 percent, whereas small standard errors in the PUF estimates lead to estimates with negligible levels of uncertainty.

Regarding future research, one could extend my analysis by estimating the SCF measurement error, with the main focus on errors arising in the survey response process as described in \citet{groves_2009_survey}. Since other sources of error are either already accounted for in the present paper or shown to be fairly marginal, computing measurement error would allow estimation of an upper bound on the SCF total standard error.

Another potentially promising avenue for future research is centered around my structural exercise. Since my analysis does not account for the impact of COVID-19, the generated projections of income shares of the top 1 percent until 2050 are largely outdated. This is the case since data-driven errors in calibration targets are most likely to be of second order of importance relative to a COVID-19 shock. Since post COVID-19 data will not be available until 2023 (the expected release date of the 2021 SCF), I plan on implementing a ``hypothetical'' COVID-19-recession shock in the model. Specifically, I intend to create a hypothetical scenario whereby the impact of COVID-19 is equivalent to or greater than that of the 2007--2009 Great Recession. Because of the severity of the ongoing pandemic and its uneven impact on US society, I expect the COVID-19 shock to have substantial and long-lasting repercussions on inequality, even when compared to those of the Great Recession.

Next, since, in its present form, the PUF does not provide any information on taxpayers' demographic and socio-economic characteristics, I plan to design an imputation procedure that would allow me to embed the SCF within the PUF. This would result in a novel data set containing as many observations as in the PUF as well as a broad range of demographic and socio-economic characteristics, which would be imputed from the SCF. With such a rich data set containing numerous potential control variables, one could answer questions on income inequality that currently cannot be addressed using the SCF or the PUF individually. For illustration, consider a problem of examining income inequality among young individuals, the working-age population, and retirees. The lack of information on age in the PUF and a small number of observations in the SCF concerning various population subgroups would make conducting such an analysis virtually impossible.

Finally, while producing better estimates of wealth inequality using the SCF would require modifying the SCF imputation procedure, introducing new interview techniques aimed at reducing item nonresponse, and/or increasing the sample size, the PUF estimates could be improved upon without changes to sampling and/or editing processes. Specifically, future research could focus on refining \citet{saez_2016_wealthinequality}'s capitalization model by allowing for more heterogeneity in the estimated rates of return or conducting a more thorough analysis of the model's vulnerability to underlying modeling assumptions.

\begin{singlespace}\small 
\bibliographystyle{aer}
\bibliography{Draft_10-20}
\end{singlespace}

\end{document}